\documentclass[a4paper,10pt, nofootinbib]{revtex4}

\usepackage{amsmath,amssymb,amsbsy,latexsym,amsthm}
\usepackage{graphicx}
\usepackage{psfrag}
\usepackage{mathrsfs}
\usepackage{setspace}
\usepackage[all]{xy}

\theoremstyle{plain}

\newcommand{\eqa}{\begin{eqnarray}}
\newcommand{\neqa}{\end{eqnarray}}
\newcommand{\be}{\begin{equation}}
\newcommand{\ee}{\end{equation}}
\newcommand{\no}{\nonumber\\}
\newcommand{\nn}{\nonumber}
\newcommand{\bea}{\begin{eqnarray}}
\newcommand{\eea}{\end{eqnarray}}

\def\R{\mathbb{R}}
\def\Z{\mathbb{Z}}
\def\C{\mathbb{C}}

\renewcommand{\1}{{\mathbb{I}}}

\def\tr{\mathrm{tr}}
\def\half{\frac{1}{2}}

\def\ra{\rangle}
\def\la{\langle}
\newcommand{\ket}[1]{|{#1}\ra}

\newcommand{\bra}[1]{\la {#1}|}

\newcommand{\SU}{\mathrm{SU}}
\newcommand{\SO}{\mathrm{SO}}
\newcommand{\SL}{\mathrm{SL}}

\newcommand{\U}{\mathrm{U}}

\newcommand{\lalg}[1]{\mathfrak{#1}}
\newcommand{\su}{\lalg{su}}

\newcommand{\Ref}[1]{(\ref{#1})}

\newcommand{\F}{{\cal F}} 

\newcommand{\hn}{\hat{n}}
\newcommand{\hp}{\flux}

\newcommand{\dd}{{\mathrm d}}

\newcommand{\coord}[1]{P_{#1}}
\newcommand{\coo}[1]{\xi_{#1}}

\usepackage{dsfont}
\newcommand{\ie}{{\it i.e.}~}

\newcommand{\id}{\mathbb{I}}
\newcommand{\flux}{{\hat{E}}}

\newcommand{\cp}{{\omega}}

\begin{document}

\begin{flushright}
AEI-2011-076
\end{flushright}

\title{Coherent states in quantum gravity: \\ a construction based on the flux representation of LQG}
\author{Daniele Oriti${}^{a}$, Roberto Pereira${}^{a,b}$ and Lorenzo Sindoni${}^{a}$ \\[.5cm]
\small \em a) Max Planck Institute for Gravitational Physics, Albert Einstein Institute, \\
\small \em Am Muehlenberg 1, 14467 Golm, Germany, EU\\
\small \em b) Instituto de Cosmologia Relatividade Astrofisica ICRA - CBPF, \\
\small \em  Rua Dr. Xavier Sigaud, 150, CEP 22290-180, Rio de Janeiro, Brazil
}

\date{\small\today}

\begin{abstract}
\noindent 
As part of a wider study of coherent states in (loop) quantum gravity, we introduce a modification to the standard construction, based on the recently introduced (non-commutative) flux representation. The resulting quantum states have some welcomed features, in particular concerning peakedness properties, when compared to other coherent states in the literature.
\end{abstract}

\maketitle\vspace{-7mm}

\section{Introduction}

Coherent states are an essential tool in the study of any quantum system, being best suited to study the correspondence with the underlying classical description of the same system, and the role of quantum fluctuations that modify it. A general review of coherent states, applied to a variety of physical phenomena can be found in \cite{klauder}. In a quantum gravity context, when a concrete definition of the (kinematical) state space of the theory is available, they are then the natural tool for testing the semi-classical limit.
Indeed, in the context of loop quantum gravity \cite{lqg}, they have been used extensively both in the canonical \cite{ashetal,gcs,BahrThiemann,stw} and in the spin foam (see, for example, \cite{sf, bianchimagliaroperini}) settings, where they define boundary states that can approximate discrete classical geometries.

\

Loop quantum gravity states are generically given by superpositions of states which have support on graphs, which can be thought of as embedded in a smooth spatial manifold or not, depending on the context\footnote{The canonical approach starts off in a continuum setting, and works with embedded graphs, even though most of the embedding, continuum information is then removed by the imposition of diffeomorphism invariance. The spin foam formulation is usually framed in a simplicial context, where the graphs used are dual to simplicial decompositions of a spatial manifold, but no embedding is assumed; often they are also interpreted as abstract, purely combinatorial graphs. These different interpretations and settings do not affect our analysis, which applies to all of them.}; each graph represents a truncation of the geometric degrees of freedom of the full, continuum theory. Looking at the classical phase space of the theory, this comes about as follows (we refer to \cite{lqg} for more details). The continuum canonical phase space is parametrized by a pair of fields $(e(x),A(x))$, representing the triad field and the (Ashtekar) connection field. In order to pass to a set of variables with nicer transformation properties under gauge transformations and easier to deal with from the analytic point of view, one then considers all possible piecewise-analytic paths $\gamma$ and 2-surfaces $S$ embedded in the spatial manifold and smeared canonical variables given by parallel transports $h_\gamma$ (path-ordered exponentials) of the connection along the paths, and fluxes $E_S$, i.e. integrals of the (dualized and densitized) triad field over the surfaces. The resulting commutation relations are very complicated in general, as one has to keep track of the intersections between paths and surfaces. Still one recognizes in them the conjugate nature of holonomies of the connection and fluxes of triad. Also, one sees that the flux variables do not (Poisson) commute, indicating a fundamental non-commutative nature of such variables already at the classical level, which should then be taken into account in the quantum theory \cite{acz}. This is the classical motivation for the non-commutative flux representation \cite{flux} for LQG quantum states we use in this paper (and used in a spin foam and GFT \cite{gft} context in \cite{aristidedaniele, aristidedaniele2}). From generic paths embedded in the spatial manifold one then passes to graphs formed by such paths and their intersection points, which allows to have a better control over the local gauge transformations acting on the canonical variables (which act, indeed, at such intersection points). To each such graph (with $N$ links) one then associates the Hilbert space, in the connection representation, $L^2(\SU(2)^{\times N}/\SU(2)^{\times V})$ defined by dividing by the action of the internal group on the nodes $V$ of the graph. The canonical structure simplifies when one considers individual links of the graphs and elementary surfaces intersecting such link at single points (so that there is a 1-1 correspondence between link $e$ and surface $S$). In this case, the canonical brackets become:
\begin{eqnarray} 
\left[ \, \hat{E}_e^i,\hat{h}_e \right] &=& i \hbar (8\pi G\gamma)  R^i \triangleright \hat{h}, \nonumber \\ \left[ \, \hat{E}_e^i,\hat{E}_{e'}^j \, \right] &=& i  \hbar (8\pi G\gamma)  \epsilon^{ij}_k \delta_{e, e'}\hat{E}_e^{k},  \nonumber \\ \left[ \, \hat{h}_{e'},\hat{h}_e \, \right] &=&\, 0, \nonumber \end{eqnarray}
where $\hbar (8\pi G\gamma) = 8\pi l_p^2 \gamma$ has the dimension of a length squared\footnote{We are using units in which $c=1$.}. One recognizes the (quantization of) the canonical brackets of the cotangent bundle of $SU(2)$, $\mathcal{T}^*\SU(2)$, for each link $e$ of the graph.

Given that we are going to work with functions and operators on functions defined on group manifolds without connecting them to physical measurements, at least at this level, it is convenient to work with dimensionless variables, reabsorbing the dimensionful quantity $ 8\pi G\gamma \hbar$ into the definition of dimensionless flux operators, which will still be denoted as $\hat{E}^i_{e}$, in such a way that the fundamental algebra reads
\begin{eqnarray} 
\left[ \, \hat{E}_e^i,\hat{h}_e \right] &=& i \cp  R^i \triangleright \hat{h}, \nonumber \\ \left[ \, \hat{E}_e^i,\hat{E}_{e'}^j \, \right] &=& i  \cp  \epsilon^{ij}_k \delta_{e, e'}\hat{E}_e^{k},  \nonumber \\ \left[ \, \hat{h}_{e'},\hat{h}_e \, \right] &=&\, 0, 
\nonumber \end{eqnarray}
where $\cp$ is an arbitrary dimensionless parameter.

\

The construction of coherent states in such setting has then as main goal the identification of kinematical states that are semi-classical in the sense that they peak, with minimal uncertainties, around classical phase space points of the continuum theory. It is then clear that one has to face several conceptual and technical issues, among which: a) the detail of the approximation of continuum geometries by data associated to graphs and thus approximating at best discrete truncation of the same continuum geometries; b) the definition of coherent states for such discrete geometries; c) the role and treatment of the sum over graphs to recover continuum configurations; d) the role of  several approximation scales in the same procedure; e) the identification of the appropriate set of observables with respect to which the approximation is best obtained; f) the construction of quantum states that are coherent with respect to such observables. The list is not exhaustive. Many such issues are discussed in \cite{ashetal,gcs,BahrThiemann,stw}. However, because of the structure of the kinematical state space of LQG, the construction of coherent states for quantum gravity necessarily has as starting point the study of suitably defined semi-classical states for the degrees of freedom associated to individual edges of the graphs on which such states are based, that is for coherent states based on a $\mathcal{T}^*\SU(2)$ phase space.

More precisely, the standard construction is based on work by Hall \cite{Hall}, who defines a coherent state transform for compact Lie groups, in analogy to the Segal--Bargmann transform for the real line. One then lifts this construction to theories of connections based on graphs by taking tensor products of Hall states, one per edge of the graph, as explained in \cite{ashetal}. The properties of those states have been extensively studied in a series of papers \cite{gcs,BahrThiemann,stw}. The essential properties are already present when restricting to a single edge, and we will focus on this case for most of this paper. We leave the next steps in the construction of proper coherent states for quantum gravity, based on complete graphs and superposition thereof, and on more interesting, and complicated observables, to future publications.

\

We start then our analysis by reviewing the construction of coherent states on a single copy of $\mathcal{T}^*\SU(2)$. As we will see, the definition of a coherent state involves two main ingredients: 1) the choice of a Gaussian on the group manifold, peaked on the origin of the phase space $\mathcal{T}^*\SU(2)$, and 2) a procedure for shifting the peak to a generic point in the same phase space, while maintaining the coherence properties. Our construction will be mostly generic on the first ingredient, even if we will refer often to the heat kernel states used by Hall in \cite{stw} as a specific example, while it will define a new procedure for what concerns the second ingredient, based on the flux representation of LQG states.
 
 \
 
In order to appreciate these ingredients and to warm up for the forthcoming discussion, let us briefly recall the simple example of a coherent state for a particle on the real line, thus with phase space $\mathbb{R}\times\mathbb{R}$.
The straightforward Gaussian 
 \be
\psi^t_{(0,0)}(x)= (2\pi t)^{-1/2}\, e^{-x^2/2t},
\ee
with $t$ the semi-classicality parameter\footnote{Throughut we use dimensionless coordinates and thus $t$ is also dimensionless.}, is a good coherent state peaking on the origin $(x_0,p_0)=(0,0)$ of phase space, and has the further nice property of having as Fourier transform a similar Gaussian in momentum representation\footnote{Note that the Fourier transform has to be defined by:
$$ \mathcal{F}(\psi)(k)= \frac{1}{\sqrt{2\pi t}} \int \dd x \, e^{ikx/t}\, \psi(x). $$}. In order to peak on a generic point in phase space one first translates the above Gaussian from $x$ to $x-q$, obtaining a coherent state peaked on  $(x_0,p_0)=(q,0)$, and finally {\it analytically continues} $q$ to $z=q-ip$. The resulting coherent state peaked on a phase space point $(q,p)$ is defined as:

\be
\psi^t_{z}(x)= (2\pi t)^{-1/2}\, e^{-(x-z)^2/2t}.
\ee

In this case, moreover, one can also understand the analytic continuation as equivalent  to a {\it translation} in momentum space with parameter $p$, or, which is the same (up to constants) as multiplying the Gaussian peaked on $(q,0)$ by a phase $e^{ix p/t}$ (the plane wave defining the Fourier transform). Indeed:
\be
 (2\pi t)^{-1/2} e^{-(x-z)^2/2t} = (2\pi t)^{-1/2} e^{(p^2 +2ipq)/2t}\; e^{-(x-q)^2/2t}e^{-ixp/t},
\ee
where the terms that do not depend on $x$ are absorbed in the normalization. 
 
It is this second way of shifting the peak of the coherent state peaked on the identity that we adopt in this paper, as an alternative to Hall's analytic continuation. 

\ 

After reviewing the standard construction (section II), we introduce the non-commutative flux representation for states on single edge (section III) , based on the group Fourier transform \cite{grouptransf}. We will be then ready to construct new types of coherent states, and to study their peakedness properties (section IV). 
 
\section{Hall states}

We start now with Hall's coherent state transform. We stress once more that this is currently the template for {\it all} coherent states defined in a quantum gravity context, if the Casimir operator is chosen as \lq complexifier\rq \cite{gcs,stw}, as we recall at the end of this section. The idea is to generalize the Segal--Bargmann transform for the real line to compact groups. 

As we have seen, one has to first define a good notion of Gaussian on the group manifold.
Hall's choice for the Gaussian is the solution to the heat kernel equation on the group:

\be
\frac{d\psi^t}{dt}=\half \Delta \psi^t,
\ee
where $\psi^t$ is a function on a Lie group $K$, s.t. $\psi^0(h)=\delta(h)$, $h\in K$, and $\Delta$ is (up to sign) the Casimir (Laplace) operator on the group. Following Peter--Weyl's theorem, the solution can be written in terms of a sum over representations as
\be
\psi^t(h)=\sum_j\, d_j\, e^{-t C_j /2}\, \chi^j(h).
\ee

$j$ labels the irreducible representations of $K$. While this is general, from now on we will restrict the discussion to the case of $K=\SU(2)$. Consequently, $j$ is half integer, $d_j=2j+1$ is the dimension of the representation $j$, $C_j=j(j+1)$ is the value of the Casimir on this representation and $\chi^j(h)$ is the character in the representation $j$. This state is peaked on the identity on the group and on the zero value for the conjugate flux variable, that is on the origin of the classical phase space. To peak outside the identity in configuration space (holonomy), one simply translates using the group multiplication:

\be
\psi^t_{h_0}(h)=\sum_j\, d_j\, e^{-t C_j /2}\, \chi^j(h h_0^{-1}).
\ee

The second ingredient is the analytic continuation; as we have seen, this is needed to peak the coherent state on a value $E_0$ for the conjugate variable (flux) different from zero. To do this consider the complexification $K^\C$ of $K$. An element $g\in K^\C$ can be written as $g=e^{iy}h$, where $h\in K$ and $y$ is in the Lie algebra of K, that we denote $\mathfrak{l}$. Parametrizing like that we see that $K^\C$ is identified with the cotangent bundle over the group $T^*(K)\sim K\times \mathfrak{l}$, and the phase space point is labelled by the pair $(h,y) \in K\times \mathfrak{l}$. The analytic continuation is then defined by the formula:

\be
\psi^t_{(h_0,y_0)}(h)=\sum_j\, d_j\, e^{-t C_j /2}\, \chi^j(h h_0^{-1}e^{-iy_0}).
\ee

The character is still the one for $K$, evaluated on analytic continued group elements. Notice also that the analytic continuation of $\SU(2)$ is the group $\SL(2,\mathbb{C})$, so that the variables $(h_0, y_0)$ define an element $H_0=h_0e^{iy_0}\in\SL(2,\mathbb{C})$ in the Cartan decomposition. This analytic continuation is proved to be unique in \cite{Hall}. This defines coherent states for a single copy of the group. 

The main properties satisfied by such states, among others \cite{Hall, gcs, stw}, which make them good candidates for being (building blocks of) quantum gravity coherent states, are the following;

\begin{enumerate}
\item {\it peakedness properties}: they peak on the appropriate points of the classical phase space:

\be
\langle \psi^t_{(h_0,y_0)} | \, \flux^i \, | \psi^t_{(h_0,y_0)}\rangle\,=\, y_0\,+\, O(t)\,\qquad \, \langle \psi^t_{(h_0,y_0)} | \, \hat{h}^i \, | \psi^t_{(h_0,y_0)}\rangle\,=\, (h_0)^i\,+\, O(t)
\ee
where $\flux$ is the flux operator (left-invariant vector field on the group manifold) and $\hat{h}$ is an appropriate operator identifying the holonomy, for example a set of coordinates on the same group manifold (we will come back to the issue of choosing such operators in the following).
\item {\it saturation of the (unquenched) Heisenberg uncertainties relations for the fundamental operators}: the states should minimize uncertainties in both the fundamental conjugate variables

\item {\it over-completeness}: the coherent states should form an over-complete basis for the Hilbert space of states; this also means that the coherent state transform from generic states to coherent states is a unitary map. 
\end{enumerate}

A comment is in order to qualify better these properties. In principle the spread  $t$ and the parameter $\cp$ appearing in the algebra of the operators we are working with are two independent parameters. However, in order
for the saturation of the Heisenberg relations to be properly satisfied with minimal fluctuations
the choice $t=\cp$ has to be made. This is the choice that we make from now on, and therefore
$t$ should not be thought as a free independent parameter, being essentially determined by
the choice of the fundamental algebra of operators.

Concerning the analytic continuation, the fact that, in the simple case of phase space $\mathbb{R}\times\mathbb{R}$, this is equivalent to a translation in momentum space suggests an alternative to the usual procedure, based on the properties of the Fourier transform of functions on $\SU(2)$. Indeed, we will take this fact as our guiding principle to introducing momentum dependence on the coherent state.
We will also show that our alternative construction leads to coherent states satisfying the same main properties listed above, and, beside aesthetic appeal and convenience in formal manipulations, improves on the standard construction by achieving a better peakedness property in the flux observables, in the sense that the new states will peak exactly on the corresponding classical flux, with not corrections of order $t$, contrary to Hall's states.

Before moving on to our new construction, let us recall briefly the relation between Hall states and other coherent states used in the quantum gravity literature. The coherent states used in the spin foam context, e.g. for calculations of lattice correlators \cite{graviton}, are Gaussians in the $\SU(2)$ representations (spin) $j$ peaked on some classical large value $j_0$ with spread $t$ (up to constant factors) and with a phase factor $e^{i j \theta_0}$, where $\theta_0$ represents the class angle of the $\SU(2)$ group element (connection) on which the state peaks. The representation parameter $j_0$ has the interpretation of the eigenvalue for the modulus of the flux (area of elementary surface associated to the link).  The dependence on the remaining components of the flux is encoded in a function of the representation $j$ and of a vector $\vec{n}\in S^2$ labeling an over-complete basis of states in the representation space corresponding to $j$. It is easy to show \cite{bianchimagliaroperini} that the Hall coherent state reduces to the states defined this way for large values of the peak area $j_0$, i.e. in a semi-classical limit. The vectors $| j , \vec{n}\rangle$ entering the same construction, in turn, are the Perelomov coherent states \cite{perelomov} for $\SU(2)$ and define a good approximation to the flux vectors, for given modulus (area), for large values of the same. As recalled in the introduction, starting from these building blocks one then constructs coherent states associated to vertices of a graph, and then to the whole graph, by tensoring coherent states associated to links and imposing  gauge invariance at the vertices of the graph. The result is the complexifier coherent states \cite{gcs,stw} if one starts from Hall states, the most general ones, in which one sees clearly the geometry behind the state and the point in the classical phase space one is peaking on, the so-called \lq coherent spin networks\rq \cite{bianchimagliaroperini}, used as said in many spin foam computations (in a simplicial setting), in turn based on the so-called Livine-Speziale coherent intertwiners \cite{eterasimone}, obtained as gauge invariant tensors of Perelomov coherent states.

\section{Non-commutative group Fourier transform and flux representation of LQG}
We now review briefly the (non-commutative) Fourier transform for the group $\SU(2)$ \cite{grouptransf}, which is the basis of the non-commutative flux representation of Loop Quantum Gravity \cite{flux}. 

The goal is to define a transform $\mathcal{F}$ mapping isometrically $L^2(\SU(2),\mu_H)$\, where $\mu_H$ is the Haar measure on the group, onto a space $L^2_\star(\R^3,\dd x)$ of functions over $\R^3\sim \su(2)$ equipped with a non-commutative $\star$-product and the Lebesgue measure. As we anticipated, the interpretation of the Lie algebra elements conjugate to the group elements is that of elementary fluxes (smeared triad fields, conjugate to holonomies of the Ashtekar connection). The first ingredient is the definition of the plane waves:

\be
e^\kappa_g: \R^3\times \SU(2) \rightarrow \U(1) \, , \;\;\; e^\kappa_g(x) := \exp\left({\frac{i}{\kappa}\coo{g}\cdot x}\right),
\ee

where $\coo{g} := (\coo{g}^{1},\coo{g}^{2},\coo{g}^{3})$ is a choice of coordinates on the group manifold and $x=x^{i}\sigma_{i}$ is a Lie algebra element. More precisely, we choose coordinates on the group as follows:

\be\label{ph}
\coo{g}^i=-\half\, \text{Tr}(|g| i \sigma^i) \, , \;\;\; |g|:=\text{sign}(\text{Tr}g)g.
\ee

Parametrizing a group element by $g=e^{i\theta \sigma\cdot \hn}=\cos\theta + i \sin\theta\sigma\cdot\hn$, one has that 

\be 
\coo{g}^i=\epsilon\sin\theta \hn^i, 
\ee

where $\epsilon=\text{sign}(\text{Tr}g)= \text{sign}(\cos \theta)$.

It is clear therefore that the definition of plane waves depends on a specific choice of coordinates on the group. The immediate consequence is that, for topological reasons, the group Fourier transform we are about to introduce will not be defined (as an invertible map) on the whole $\SU(2)$, as it is impossible to cover the whole group manifold by a single coordinate patch. We will discuss below how to overcome this limitation.
The plane wave is labelled by a parameter $\kappa$ that will be related, later on, to the spread of the semiclassical states, denoted by $t$. Indeed, from the algebra of fundamental operators in LQG, one can already expect $t = \kappa$. We will provide below a motivation for this identification. In order not to overload notations we will anyway leave it implicit for most of the equations. 

Some relations for the plane waves will be useful later on: 

\be
\overline{e_h(x)}=e_h(-x)=e_{h^{-1}}(x).
\ee

The group Fourier transform is then defined as:

\be
\mathcal{F}(f)(x):= \int \dd g \, e_g(x)\, f(g). 
\ee

Notice that the plane waves so-defined do not distinguish between $g$ and $-g$. As a consequence, the above Fourier transform is not invertible for functions on $\SU(2)$, but only for functions on $\SU(2)$ that are also invariant under the same discrete symmetry. These can be identified with functions on $\SO(3)\!\simeq\! \SU(2)/\Z_2$, for which the above is a proper Fourier transform.

The definition of the plane wave induces a natural algebra structure on the image of the Fourier transform. The product is defined on plane waves by:

\be
e_{g_1}\star e_{g_2} := e_{g_1 g_2}.
\ee

The scalar product in $L^2_\star(\R^3,\dd x)$ is defined by:

\be
\langle u,v \rangle_\star := \frac{1}{8\pi}\int \dd x^3 (\bar{u}\star v)(x),
\ee
and the inverse Fourier transform is given by:

\be
f(g)=\frac{1}{8\pi}\int\dd x^3 (\F(f)\star e_{g^{-1}})(x).
\ee

Normalizations are for $\SO(3)$, in which case $\epsilon$ is always equal to one. One can show \cite{flux} that with this scalar product the Fourier transform defines an unitary map between the spaces $L^2_\star(\R^3,\dd x)$ and $L^2(\SO(3))$. The same can be extended to generic cylindrical functions associated to arbitrary graph embedded in a 3-manifold, and thus to the whole kinematical Hilbert space of LQG (restricted to $\SO(3)$), in a way that, moreover, preserves cylindrical consistency requirements \cite{flux}.

\ 

The restriction to $\SO(3)$ is somewhat unsatisfactory for some applications, and it is useful to lift it, especially for a proper comparison between our construction and the usual coherent states previously defined in the literature, in particular Hall coherent states, which use the full $\SU(2)$ manifold. This generalization can be achieved in more than one way \cite{majidfreidel, karim, maiteflorianetera}. We describe here, briefly, the extension defined in \cite{karim}, to which we refer for more details.

Given that main obstruction to a 1-1 map between $\SU(2)$ and $\mathbb{R}^3$ is topological, we define three subsets of $\SU(2)\simeq S^3$, corresponding to its northern hemisphere $U_+$, southern hemisphere $U_-$ and equator $U_0$: $\U_\epsilon = \{ g(\vec{P}_g,  P^0_g) \in \SU(2) | \text{sign}(P_0) = \epsilon \}$, with $\vec{P}_g = \epsilon \vec{\xi}_g = \sin\theta \vec{n}$, $P_g^0 = \cos\theta$ and $\text{sign}(0)=0$ (this is the standard coordinate system on the 3-sphere embedded in  $\mathbb{R}^4$, with $(\vec{P}, P_0)$ such that $P^2_0+P^2_1+P^2_2+P^2_3 = 1$). In other words, we decompose the space of generalised functions on $\SU(2)$ into subspaces:

$$
C(\SU(2))^* \simeq C(U_+)^* \oplus C(U_-)^* \oplus C(U_0) 
$$

so that, for any  $f\in C(\SU(2))^*$, $f\, =\, f_+\, \oplus\, f_-\,\oplus\,f_0$, with $f_{\pm,0} \, =\, f\, I_{\pm,0}$, where $I_{\pm,0}$ are characteristic functions on $U_{\pm,0}$.
Clearly, the elements of $C(U_0)$ have necessarily distributional nature (with respect to the Haar measure). Therefore, ordinary functions on $\SU(2)$ have only components in $C(U_{\pm})$.
The decomposition is characterized by projections that can be associated to polarization vectors: 

$$
\xi_+ = 1 \oplus 0 \oplus 0\; , \qquad \xi_- = 0 \oplus 1 \oplus 0 \;\; , \qquad  \xi_0 = 0 \oplus 0 \oplus 1
$$

The Fourier transform $\mathcal{F}$ that is bijective on the full $C(\SU(2))^*$ and respects the above decomposition is then defined in terms of the plane waves (generalizing the above ones):
\be
e^\kappa_{g,\epsilon}: \R^3\times \SU(2) \rightarrow \U(1) \, , \;\;\; e^\kappa_{g,\epsilon}(x) := \exp\left({\frac{i}{\kappa}P_{g}\cdot x}\right)\, \xi_\epsilon\;\;\; ,
\ee

and maps generalised functions $f = f_+ \, \oplus \, f_- \,\oplus\, f_0$ on $\SU(2)$ into a multiplet of functions $\mathcal{F}(f) = \tilde{f} = \tilde{f}_+ \,\oplus\, \tilde{f}_- \,\oplus\, \tilde{f}_0$ on $\mathfrak{su}(2)\simeq\mathbb{R}^3$, which can be denoted $C_\kappa(\mathfrak{su}(2))$. 

For ordinary functions on $\SU(2)$ it looks simply as:
\eqa
&&\tilde{f}_\pm(x)\, \xi_\pm\,=\, \int_{U_\pm} dg \, f_\pm(g) \, \exp\left({\frac{i}{\kappa}P_{g}\cdot x}\right)\, \xi_\pm = \no 
&& =\, \int d^3\vec{P}_g \frac{1}{2\sqrt{1 - |\vec{P}_g^2|}} f\left(\vec{P}_g, \pm\sqrt{1-|\vec{P}_g|^2}\right)\, \exp\left({\frac{i}{\kappa}P_{g}\cdot x}\right) \,\xi_\pm.
\neqa

{ The function $\tilde{f}_0$ can be obtained similarly by integration over the equator of $SU(2)$, as discussed in \cite{karim}.}

This map from $C(\SU(2))^*$ and $C(\mathbb{R}^3)_\kappa^*$ is invertible and the inverse map can be expressed using a $\star$-product \cite{karim}.

The $\star$-product between plane waves $e_{g}(x)$, inducing a product on general elements of $C(\SU(2))^*$ by linearity, takes into account the decomposition of the domain and target spaces of the Fourier transform, and is defined as:
\eqa
&&\left(e_{g_1,\epsilon_1}\,\star\,e_{g_2,\epsilon_2}\right)(x)\,=\,\exp\left({\frac{i}{\kappa}P_{g_1}\cdot x}\right)\, \xi_{\epsilon_{1}} \star \,\exp\left({\frac{i}{\kappa}P_{g_2}\cdot x}\right)\, \xi_{\epsilon_{1}} \, \equiv \no
&&\equiv \, \exp\left({\frac{i}{\kappa}P_{g_1g_2}\cdot x}\right)\, \xi_{\epsilon_{12}}\,=\,e_{g_1g_2,\epsilon_{12}}(x)
\neqa
where explicitly \cite{karim}:
\bea
\vec{P}_{g_1g_2}\, &=&\, \epsilon_{2}\sqrt{1 - (\kappa| \vec{P}_{g_2}|)^2} \vec{P}_{g_1}\,+\,\epsilon_{1}\sqrt{1 - (\kappa| \vec{P}_{g_1}|)^2} \vec{P}_{g_2}\, + \, \kappa\, \vec{P}_{g_1} \wedge \vec{P}_{g_2} \\  \epsilon_{12} &=& \text{sign}\left( \epsilon_{1}\epsilon_{2}\sqrt{1 - (\kappa| \vec{P}_{g_1}|)^2}\sqrt{1 - (\kappa| \vec{P}_{g_2}|)^2} \, - \, \kappa\, \vec{P}_{g_1} \cdot \vec{P}_{g_2}\right). \eea

The $\star$-product between two arbitrary functions $\Phi_{1,2} = \mathcal{F}(\phi_{1,2})$ is defined to be dual to the convolution product $\circ$ for functions on the group, and then implicitly defined by the formula:

\be
\Phi_1 \star \Phi_2 \, =\, \mathcal{F}\left( \mathcal{F}^{-1} (\Phi_1)\, \circ \, \mathcal{F}^{-1} (\Phi_2)\right),
\ee

which can in turn be expressed as a nonlocal integral as \cite{karim}:

\be
\left( \Phi_1 \star \Phi_2\right) (x) \, =\, \sum_{\epsilon_{1}, \epsilon_2}\, \int_{\mathbb{R}^3} dy dz\, K_{\epsilon_1\epsilon_2}(x,y,z)\, \star_{y,z} \Phi_{1\epsilon_1}(y)\Phi_{2\epsilon_2}(z),
\ee

with 

$$
K_{\epsilon_1\epsilon_2}(x,y,z)\,=\, \int_{\SU(2)}dg_1 \,dg_2\, e^{i\left( P(g_1)\cdot y\, +\, P(g_2)\cdot z\, +\, P(g_1g_2)\cdot x\right)}\, \xi_{\epsilon_{12}}.
$$

Another useful property, following from the above definition, is that 
\be
\int dx\, \left( \Phi_1 \star \Phi_2\right) (x) \, = \int dx \left( \Phi_{1+}\xi_+ \star \Phi_{2+}\xi_+\, + \,\Phi_{1-}\xi_- \star \Phi_{2-}\xi_- \right) (x)\, =\, \int \dd g\, \left( \Phi_1 \, \Phi_2\right) (g).
\ee
More details on this $\star$-product for generic functions can be found in \cite{karim}.
One main feature of this definition of Fourier transform on $\SU(2)$ is that it is covariant under the standard linear action of $D\SU(2)$, the Drinfeld double, a quantum group deformation of the 3d euclidean Poincar\'e group. In particular, this allows a nice definition of (non-commutative) translation in the $\mathfrak{su}(2)$ algebra for $C(\mathbb{R}^3)$, a feature that we are going to exploit in the following. This is the main advantage of this particular definition, with respect to other proposals in the literature \cite{majidfreidel, maiteflorianetera}, beside the role it plays in the applications to spin foam models and group field theories \cite{aristidedaniele, aristidedaniele2, aristidefloriandaniele}. While the split into subregions of $\SU(2)$, and thus the use of a multiplet structure for the target functions,  is quite natural from the topological point of view, one may still want, instead of such multiplets, a unique function on $\mathbb{R}^3$ as a target. This is what, for example, the definition of \cite{maiteflorianetera} achieves. 

In any case, using this definition of Fourier transform, the split of the $\SU(2)$ manifold into subspaces is needed in order to define, via the above procedure, an invertible Fourier map, at the cost of complicating slightly the notation (with polarization vectors, characteristic functions etc). Therefore, in the following we will make use of it whenever calculations are performed and reported in the non-commutative Fourier space. For standard manipulations of functions on $\SU(2)$, instead, we will stick to the usual, simpler notation.

\section{A modified construction of coherent states on $\SU(2)$}
We now proceed, using also the flux (Lie algebra) representation introduced above, to give a modified definition of coherent states peaked on an arbitrary point of the phase space $\mathcal{T}^*\SU(2) \simeq \mathfrak{su}(2) \times \SU(2)$. The starting point is a Gaussian state on $\SU(2)$ that is peaked on the origin $(p_0,h_0) = (0,e)$.

One good choice is, as for the Hall state, the heat kernel on $\SU(2)$, that we already introduced, decomposed in representations as:
$$
\psi^t(h)=\sum_j\, d_j\, e^{-t C_j /2}\, \chi^j(h).
$$
Its Gaussian form in group space is made apparent by its expression in coordinates on $\SU(2)$:
\be
\psi^t(h) = \frac{-1}{(4 \pi t)^{\frac{3}{2}}} \sum_{n=-\infty}^{\infty} \bigg ( \frac{\theta(h) + 4 \pi n}{2\sin(\frac{\theta(h)}{2})} \exp \Big [ -\,\frac{1}{2t} \big( \theta(h) + 4 \pi n \big)^2 - \frac{t}{8} \Big ] \bigg )   .
\ee
where the sum is over geodesics over the group manifold, connecting the relevant point to the \lq north pole\rq ~(group identity), and is required to ensure the correct periodicity \cite{camporesi}. 

Using the noncommutative Fourier transform introduced in the previous section, one can make apparent also its Gaussian form in Lie algebra (flux) space. 
To show this, let us use the mentioned split of the $\SU(2)$ manifold into upper and lower hemispheres (each isomorphic to the $\SO(3)$ group manifold: $\psi^t = \psi^t_+ \xi_+ \, +\, \psi^t_- \xi_-$. The restriction $\psi^t_{\pm}(h)$ of the heat kernel to each such hemisphere gives then the heat kernel on  $\SO(3)$, also obtained from the one on $\SU(2)$ as $\psi^t_+(\theta) = \psi^t(\theta) + \psi^t(2\pi - \theta)$ \cite{camporesi}. 
The Fourier transform of this function is then computed as  
\be
\mathcal{F}[\psi_t](x) \, =\, \sum_\epsilon \mathcal{F}[\psi_{t,\epsilon}](x)\, \xi_\epsilon\,=\, \sum_\epsilon \int_{\SU(2)} dg\, \psi^t_\epsilon(g)\, e^{\frac{i}{\kappa} P_g \cdot x}\, \xi_{\epsilon}    \qquad .
\ee
A simple calculation (see \cite{majidfreidel}\footnote{In the same paper it is also shown that the same result is obtained by solving directly the Lie algebra expression for the heat kernel equation, which one can also see as the {\it definition} of a Gaussian function on Lie algebra space.}) shows that the resulting function on the algebra is:

\be
\mathcal{F}[\psi_{t,\epsilon}](x)\,=\, e_\star^{- \frac{t}{4 \kappa^2}\, x^{\star 2}}\, =\, e_\star^{- \frac{t}{4 \kappa^2}\, x\star x}
\ee
where the non-commutative exponential is defined (see also \cite{danielematti}) by power series expansion into $\star$-products of the coordinate functions, and the dependence on $t$ and $\kappa$ suggests to identify the two (possibly up to constants), as we will in fact motivate further in the following. We see, therefore, that the heat kernel on the group is indeed the obvious Gaussian in Lie algebra space, centered again on the phase space point $(x_0,h_0)=(0,e)$. 

\

Another possible Gaussian, defined in \cite{majidfreidel}, that would lead to a different definition of coherent states, and that should be kept in mind for an alternative concrete implementation of our construction, is given by:

\be
g^t_{\1}(h):= \exp\left(\frac{2}{t}\chi(h)\right).
\ee

\

Having defined a suitable Gaussian state centered on the origin of the classical phase space, the task becomes that of defining from it a new state centered around an arbitrary phase space point. In order to center it around an arbitrary value of the classical holonomy, one can use the translation on the group manifold, as in the standard LQG definition of coherent states, and as it is suggested by the very definition of the classical phase space (see also \cite{danielematti}). 

One gets then again to the state: 
\be
\psi^t_{h_0}(h)= \psi^t(h h_0^{-1})\,=\,\sum_j\, d_j\, e^{-t C_j /2}\, \chi^j(h h_0^{-1}).
\ee

{ The expression of the same state in Fourier space further elucidates the relation between the translations on the group manifold and the plane waves:}

\bea
\tilde{\psi}^t_{h_0}(x)\,&=&\,\sum_\epsilon\,\tilde{\psi}^t_{h_0,\epsilon}(x)\,\xi_\epsilon\,=\,\sum_\epsilon\,\mathcal{F}\left[\psi^t_{h_0}\,I_\epsilon\right](x)\,\xi_\epsilon\,= \, \sum_\epsilon \int dh\,  \left( \psi^t( \, \cdot \,h_0^{-1}) \, I_{\epsilon}(\cdot)\,\xi_{\epsilon(\cdot)}\right)(h)\, e^{\frac{i}{\kappa}\, P(h)\cdot x}\,=\nn \\
&=& \,  \sum_\epsilon \int dh\,  \left( \psi^t( \,\cdot\,h_0^{-1}) \, I_{\epsilon}(\cdot)\,\xi_{\epsilon(\cdot)}\right)(hh_0)\, e^{\frac{i}{\kappa}\, P(hh_0)\cdot x}\,= \no 
&=&\,  \sum_\epsilon \int dh\,  \psi^t(h ) \, I_{\epsilon}(hh_0)\,\xi_{\epsilon(hh_0)}\, e^{\frac{i}{\kappa}\, P(hh_0)\cdot x}\,=\, \nn \\
&=& \, \sum_{\epsilon} \int dh \, \psi^t(h)\, I_{\epsilon}(hh_0)\,e^{\frac{i}{\kappa}\,P(h)\cdot x}\xi_{\epsilon(h)}\,\star\, e^{\frac{i}{\kappa}\, P(h_0)\cdot x}\,\xi_{\epsilon(h_0)}\,=\, \sum_\epsilon\,\left(  \psi^t \, \star\, e^\kappa_{h_0,\epsilon_0} \right)_\epsilon(x)\,\xi_{\epsilon} \,= \,\nn \\ &=&\,\left(  \psi^t \, \star\, e^\kappa_{h_0} \right)(x)\, =\, \psi^t(x)\,\star\,e^{\frac{i}{\kappa}\,P(h_0)\cdot x}.
\eea

Indeed, it is given by the $\star$-multiplication of the state centered in $(0,e)$, which we have seen to be the natural Gaussian state at the origin, by the plane wave (phase) corresponding to the argument (conjugate variable) $h_0$.

Apart from its re-expression in non-commutative flux variables, if one chooses the heat kernel on the group manifold as an initial Gaussian, the above state is the usual (complexifier) coherent state used in LQG \cite{gcs,stw}, for zero flux. 
The next task is to shift the location of the peak of the coherent state from the origin of the Lie algebra (flux) coordinates to a generic flux. As we have discussed, this step is achieved in the usual construction by analytic continuation of the peak group element $h_0$ from $\SU(2)$ to $\SL(2,\mathbb{C})$.  

The alternative we propose is the one that is naturally suggested by the non-commutative flux representation itself, and amounts to performing a simple translation in Lie algebra space. This is achieved by multiplying the original state, expressed in group variables, by a plane wave with Lie algebra argument $-x_0$, if $x_0$ is the value on the algebra where we want our new state to be peaked.  

Denoting $\psi^t_{(h_0,x_0)}(h):=\psi^t_{h_0}(h) e_h(-x_0)$ and going to Fourier space, one has:

\eqa
&&\F(\psi^t_{(h_0,x_0)})(x) = \sum_\epsilon\,\int \dd h \, e_h(x) e_h(-x_0) \left(\psi^t_{h_0} I_\epsilon\right)(h)\, \xi_{\epsilon} = \no
&& = \sum_\epsilon \int \dd h \, e_h(x-x_0) \psi^t_{h_0}(h) I_\epsilon(h) \xi_\epsilon\, = \F(\psi^t_{h_0})(x-x_0).
\neqa

The same state can be written, with a certain, suggestive abuse of notation, as:

\be
\F(\psi^t_{(h_0,x_0)})(x)\,\propto\, e_*^{- \frac{t}{4 \kappa^2}\, (x\,-\,x_0)^2}\,\star \, e^{\frac{i}{\kappa}\,P(h_0)\cdot x}
\ee
indicating once more its natural construction as a coherent state on phase space (as it is obvious, we have dropped, in the above formula, a factor function of $h_0, x_0$ and $\kappa$, to highlight the dependence on $x$ only).

We have thus shown that our general procedure produces at least one very reasonable candidate for a coherent state on a single edge of the group, very closely related to, but still different from, the standard complexifier coherent state of Thiemann and collaborators. We will now analyze the general properties of the coherent states so constructed. Some of them would follow quite generically from the construction itself, and will not depend on specific examples. Others would instead rely on specific choices. When a specific choice has to be made to carry out the calculation, we will choose the one illustrated in detail above.

\subsection{Expectation values}
Let us now compute the expectation value of the flux operator on this state. Note that we are using dimensionless operators.The action of the flux operator on group space is given by:

\be
\flux^i \triangleright f(g) = i t R^i \triangleright f(g) = i t \left. \frac{d}{ds} f(e^{is\sigma^i}g)\right|_{s=0} .
\ee
The $t$ comes from the commutator:
\be
[ \flux^i,\hat{h} ]  = i t R^i \triangleright \hat{h}.
\ee

On Fourier space this action is given by:
\begin{eqnarray}
&&\F(\flux^i \triangleright f)(x) = \F\left(\sum_\epsilon(\flux^i \triangleright f)_\epsilon\right)(x)\, \xi_\epsilon = it \sum_\epsilon \int\dd g \,  \left. \frac{d}{ds} f(e^{is\sigma^i}g)\, I_\epsilon(g)\, e_g(x) \, \xi_{\epsilon(g)} \right|_{s=0} \, = \nn \\
&&=i t \sum_\epsilon \left. \frac{d}{ds} \int\dd g \, I_\epsilon(e^{-is\sigma^i}g)\, e_{e^{-is\sigma^i}g}(x) \xi_{\epsilon(e^{-is\sigma^i}g)} \, f(g) \right|_{s=0} = \no
&&= it \sum_\epsilon \int\dd g \,  \left. I_\epsilon(e^{-is\sigma^i}g) \, \frac{d}{ds} e_{e^{-is\sigma^i}g}(x) \xi_{\epsilon(e^{-is\sigma^i}g)}\, \right|_{s=0} f(g)  = \no
&&= it \sum_\epsilon\, \int\dd g \,  \left. I_\epsilon(e^{-is\sigma^i}g)\, \frac{d}{ds} e_{e^{-is\sigma^i}}(x)\,\xi_{\epsilon(e^{-is\sigma^i})}\right|_{s=0}\star e_{g}(x) \, \xi_{\epsilon(g)}\, f(g) = \no
&&=i \frac{t}{\kappa}\sum_\epsilon\,\int\dd g \, I_\epsilon(g)\, (-ix^i) \star e_{g}(x) f(g) \xi_{\epsilon(g)}\, = \no
&&= \frac{t}{\kappa} \sum_\epsilon\, x^i \star \F(f_\epsilon)(x)\, \xi_\epsilon\, = \frac{t}{\kappa} x^i \star \F(f)(x), 
\end{eqnarray}
which justifies the choice $\kappa=t$, which we make from now on.

The expectation value is then computed on Fourier space\footnote{Recall that $\F(f)(x)$ is always intended to be defined as $\F(f)(x)= \sum_\epsilon \F(f_\epsilon)(x) \xi_\epsilon$, and that the integral of the $\star$-product of two functions is a sum over the integrals of the product of their positive and negative components, with no mixed terms, as shown in the previous section.}:
\eqa
&&\bra{\psi^t_{(h_0,y_0)}} \flux^i \ket{\psi^t_{(h_0,y_0)}} =  \int\dd x \overline{\F(\psi^t_{(h_0,y_0)})}(x) \star x^i \star \F(\psi^t_{(h_0,y_0)})(x) = \no
&&=  \int\dd x \overline{\F(\psi^t_{h_0})}(x-y_0) \star x^i \star \F(\psi^t_{h_0})(x-y_0) = \no
&&=  \int\dd x \overline{\F(\psi^t_{h_0})}(x) \star (x+y_0)^i \star \F(\psi^t_{h_0})(x) = \no 
&&= y_0^i ||\psi^t_{h_0}||^2 + \int\dd x \overline{\F(\psi^t_{h_0})}(x) \star x^i \star \F(\psi^t_{h_0})(x) =\no
&&= y_0^i ||\psi^t_{h_0}||^2 + \bra{\psi^t_{h_0}} \flux^i \ket{\psi^t_{h_0}}.
\neqa

Remarking that  $||\psi^t_{h_0}||^2=||\psi^t_{(h_0,y_0)}||^2$, we see that the condition for the expectation value being equal to the classical value $y_0$ is that it is zero for the original Gaussian. Let us see what properties of the Gaussian ensure that this condition is satisfied. From the already assumed property $\psi^t_{h_0}(h)=\psi^t_{\1}(hh_0^{-1})$ it follows that, since the action of $\flux^i$ is right invariant, it commutes with right translation on the group and the condition on the expectation value can be asked for the state peaked on the identity. 

\bea
&&\bra{\psi^t_{(h_0,0)}} \flux^i \ket{\psi^t_{(h_0,0)}} = \int dh\, \overline{\psi^t_{(e,0)}}(hh_0^{-1})\,\flux^i\,\psi^t_{(e,0)}(hh_0^{-1}) \nonumber \\ 
&&=\int dh\, \overline{\psi^t_{(e,0)}}(h)\,\flux^i\,\psi^t_{(e,0)}(h)  = \bra{\psi^t_{(e,0)}} \flux^i \ket{\psi^t_{(e,0)}}.
\eea

Finally, assuming that this state is a class function, which implies it is a function on the characters on the group, finally implies that the expectation value of $\flux^i$ is zero. 

Those conditions are met, for example, by the heat kernel used to define the Hall states, as well as by the alternative Gaussian state mentioned in the previous section. Once more, we will keep the discussion as general as possible, assuming the conditions described above, and restrict to a specific Gaussian only when necessary. 

As a welcome result, these states are then peaked {\it exactly} on the classical value of the flux that is used to define the state itself.
 
 \
 
The next task is to compute the expectation value, in our coherent states, of the conjugate operator to the flux. As  stated, this is an ill-posed question, because there is no operator defined in the kinematical Hilbert space of LQG, whose commutation relations with the flux are exactly canonical. The natural conjugate operators are however the holonomies of the Ashtekar connection along the same link of the graph, whose commutator with the flux is proportional to the holonomy itself. Still, the question remains to be better defined, as one has to choose a specific function on the group to represent the holonomy at the quantum level. Any such function will act by multiplication in the connection representation and by (generalized) translation in the flux representation \cite{flux}. The usual choice is to consider characters of the group element representing the holonomy, in the fundamental representation. Here we make a different choice and consider instead coordinate functions on the group manifold, as the appropriate operators to be used to represent holonomies at the quantum level.
We need first to choose a good coordinate system for the group manifold. A natural one is given by equation \Ref{ph}. { As we have already pointed out, since those coordinates are related to $SO(3)$ and will be insensitive to the fact that the configuration space is indeed $SU(2)$, it will be more convenient to use instead the $P$s defined as}
\begin{equation}
P^{i}(h) = -\frac{i}{2} \tr(h\sigma^i) = \sin\theta \, n^i,
\label{eq:defp}
\end{equation}
with the label of the corresponding hemisphere being controlled by
\begin{equation}
P_0(h) = \cos{\theta}.
\end{equation}
In this way, all the integrations can be straightforwardly performed on $SU(2)$.

For the following calculations, it is useful to remind the composition formula for the coordinates
\begin{equation}
\coord{hh_{0}}^{i} = \cos{\theta_{0}} \coord{h}^{i} + \cos{\theta} \coord{h_{0}} ^{i} -  \epsilon_{ijk} \coord{h}^{j}\coord{h_{0}}^{k}.
\end{equation}

Let us then compute $\la \hat{\coord{h}^i} \ra$ for a normalized state:
\eqa
&&\bra{\psi^t_{(h_0,y_0)}} \hat{\coord{h}^i} \ket{\psi^t_{(h_0,y_0)}} =  \int\dd h \overline{\psi^t_{(h_0,y_0)}(h)} \coord{h}^i \psi^t_{(h_0,y_0)}(h)= \no
&& = \int\dd h \overline{\psi^t_{h_0}(h)} \coord{h}^i \psi^t_{h_0}(h) = \int\dd h \overline{\psi^t_{\1}(h)} \coord{(hh_0)}^i \psi^t_{\1}(h) = \no
&& = \int\dd h |\psi^t_{\1}(h)|^2 \coord{(hh_0)}^i = \int\dd h |\psi^t_{\1}(h)|^2 (\coord{h} \oplus \coord{h_0})^i = \no
&& =  \int\dd h |\psi^t_{\1}(h)|^2 \left(\cos(\theta) \coord{h_0} + \cos{\theta_0} \coord{h} - \coord{h}\wedge \coord{h_0}\right)^i \no
&& =  \left(\int\dd h |\psi^t_{\1}(h)|^2\cos{\theta}\right) \coord{h_{0}}^i. 
\neqa
Here we are using the fact that, due to the {symmetry} properties\footnote{The result can be obtained observing that, being the heat kernel a class function, the expectation value
\begin{equation*}
v^i=\int dh |\psi(h)|^2 P^i(h) = \int dh |\psi(g^{-1}hg)|^2 P^i(h) = \int dh' |\psi(h')|^2 P^i(ghg^{-1})=
R^{i}_{j}(g) v^j
\end{equation*}
is an  invariant vector under the adjoint action of the group (that is, due to the definition \eqref{eq:defp}, $SO(3)$ rotations), and hence it must be the null vector.} of the heat kernel (and as an explicit integration can easily show), 
\begin{equation}
\int dh |\psi^t_{\1}(h)|^2 \coord{h}^i = 0.
\end{equation}

The net result is that
\be
\la \coord{h}^{i}\ra = a(t) \coord{h_{0}}^{i}, \qquad -1<a(t) = \langle \cos(\theta) \rangle = \langle \coord{h}^0 \rangle< 1,
\ee
as it follows from analyticity properties of the heat kernel (we will describe the behavior of 
$a(t)$ when dealing with the fluctuations).

Therefore, the expectation value
cannot be exactly equal to the classical value unless $h_{0}=\mathbb{I}$, even though one can make it arbitrarily close to it, by making the semiclassical parameter $t$ small enough. To correct this behavior, in particular the fact that the expectation value in the state $\psi^t_{h_0}$ depends on the parameter $t$, one needs to use different coordinates on the group.

It is instructive to see how generic is this result. In evaluating the expectation value of a certain
function of the holonomy, $\mathscr{O}(h)$, that we assume to be regular enough to admit a Peter--Weyl decomposition in representations
\begin{equation}
\mathscr{O}(h) = \sum_{j} d_j (O_{j})^{a}_b (D^j(h))_{a}^b,
\end{equation}
the result is, straightforwardly
\bea
&&\la \mathscr{O}(h) \ra_{h_0}=
\sum_{j}d_j (O_{j})^{a}_b \int dh |\psi^t_{\1}(h)|^2(D^j(hh_0))_{a}^b = \sum_{j}d_j (O_{j})^{a}_b (T^j)_{c}^b (D^{j}(h_0))^{c}_a\, ,
\nonumber
\\
&&(T^{j})_{a}^b=\int dh |\psi^t_{\1}(h)|^2  (D^j(h))_{a}^b \,. 
\eea
Now, if the Gaussian chosen to build the state is a class function (and hence more general than the heat kernel), this integral
is left invariant by the adjoint action of the group $SU(2)$:
\begin{equation}
(\tilde{T}^{j})_{d}^c= (D_{j}^\dagger(g))^{c}_b(T^{j})_{a}^b(D_{j}(g))^{a}_d=\int dh |\psi^t_{\1}(h)|^2 (D^j(g^{\dagger}h g))_{d}^c = ({T}^{j})_{d}^c,
\end{equation}
which implies that
\begin{equation}
(T^{j})^{a}_{b} = \mathscr{I}_{j}(t) \delta^{a}_{b} \, ,\qquad
\mathscr{I}_{j}(t) = \frac{1}{d_j} \int dh |\psi^t_{\1}(h)|^2  \chi^j(h) = \frac{1}{d_j}\la \chi_{j}(h)\ra\,.
\end{equation}

Therefore, the expectation value reads:
\begin{equation}
\la \mathscr{O}(h) \ra_{h_0} =\sum_{j}d_j\mathscr{I}_j(t) (O_{j})^{a}_b (D^{j}(h_0))^{b}_a = \tilde{\mathscr{O}}_t(h_0),
\end{equation}
which is stating that the expectation value is a function of the position of the peak in the $SU(2)$ configuration space, but that it is a different function, which, in addition, depends on $t$. This should not come as a surprise. Indeed, for the particular Gaussians given by the heat kernel coherent state \cite{gcs}, it has been shown that the state is an eigenstate of an annihilation operator
\begin{equation}
\hat{A} = \exp\left(+\frac{t}{2}\triangle\right) h \exp\left(-\frac{t}{2}\triangle\right),
\end{equation}
which is a nonpolynomial function of the holonomies and fluxes. Therefore, there is no guarantee that the expectation values of holonomy operators do match the value of the operator evaluated on the peak, a result that holds only for polynomial functions of the creation and annihilation operators.

However, it might be possible to introduce { operators (functions of the holonomy) whose expectation values are given by the corresponding classical phase space function the $SU(2)$ element specifying the peak.} To show this, it is convenient to consider functions of the coordinates $\coord{h}^i$ that we have introduced before. We will try to find operators such that
\begin{equation}
\la \varphi^i(h) \ra_{h_0} \approx \varphi^i(h_0) + O(\coord{h_0}^2), \qquad |\coord{h_0}| \ll 1
\end{equation}
which is the kinematical regime in which the holonomies are close to the identity matrix. Physically, this regime might correspond to a low curvature region (or to a very fine grained decomposition of a generically curved space).

We start from an ansatz
\begin{equation}
\varphi^i(\coord{h})= f(\coord{h}^2) \coord{h}^i \label{newcoord}
\end{equation}
The form is completely dictated by the tensorial structure: we do not have enough vectors to construct anything else.

Under composition:
\begin{equation}
\coord{h} \rightarrow \cos(\theta_0)\coord{h} + \cos{\theta} \coord{h_0} - (\coord{h}\wedge\coord{h_0}).
\end{equation}
Now, working at first order in $\theta_0$,
\begin{equation}
\coord{h} \rightarrow \coord{h} + \cos{\theta} \coord{h_0} - (\coord{h}\wedge\coord{h_0}),
\end{equation}
whence
\bea
&&\int dh |\psi^t_{h_0}(h)|^2 f(\coord{h}^2)\coord{h}^i \approx \int dh|\psi^t_{\mathbb{I}}(h)|^2 \left( f(\coord{h}^2)\coord{h}^i+ f(\coord{h}^2) [\cos{\theta}\coord{h_0}^i - (\coord{h}\wedge\coord{h_0}))^i]
\right. \nonumber \\
&&\left.
  + 2 f'(\coord{h}^2) \coord{h}^i \coord{h}^j [\cos{\theta} \coord{h_0}^j - (\coord{h}\wedge\coord{h_0})^j] \right).
\eea
Expanding the products, the first and third terms average to zero, the second gives a contribution proportional to $\coord{h_0}$, as the fourth, though the tensorial structure is a bit less trivial. The fifth term is obviously identically zero.
Define the following integrals
\bea
&&\int dh|\psi^t_{\mathbb{I}}(h)|^2 f(\coord{h}^2)\cos{\theta} = J_1(t)
\\
&&2\int dh|\psi^t_{\mathbb{I}}(h)|^2  f'(\coord{h}^2) \coord{h}^i \coord{h}^j\cos{\theta} = J_2(t) \delta^{ij}
\\
&&J_2 = \frac{2}{3}\int dh|\psi^t_{\mathbb{I}}(h)|^2  f'(\coord{h}^2) \coord{h}^2\cos{\theta}
\eea
These are the only integrals contributing to the result, at this order in the expansion in $\theta_0 \approx |\coord{h_0}|$. 
At first order in $\theta_0$ we get:
\begin{equation}
\int dh |\psi^t_{h_0}(h)|^2 f(\coord{h}^2) \coord{h}^i \approx 0 + (J_1(t)+J_2(t)) \coord{h_0}^i 
\end{equation}
and we want it to be equal to
\begin{equation}
f(\coord{h_0}^2) \coord{h_0} \approx f(0)\coord{h_0}
\end{equation}
at this order, and with $f(0)$ independent from $t$.
This implies, as a first condition, that
\begin{equation}
\frac{d}{dt}(J_1(t)+J_2(t)) = 0,
\end{equation}
which is a complicated integro-differential equation for $f$. While giving a necessary and sufficient condition for $f$ to be a solution of this equation clearly stands beyond the scope of this paper, we can provide a sufficient condition of the form
\begin{equation}
f(x) + \frac{2}{3} xf'(x) = \frac{\lambda}{\sqrt{1-x}},
\end{equation}
which comes from the requirement that the integrand of $J_1+J_2$ is just  the modulus square of the (normalized!) Gaussian, up to an arbitrary constant, $\lambda$ (independent from $t$).
The only solution to this differential equation regular at $x=0$ is
\begin{equation}
f(x) = 3\lambda \left( \frac{  \arcsin(\sqrt{x})-\sqrt{x(1-x)}}{2 x^{3/2}} \right) \label{sol}
\end{equation}
Taylor-expanding around the origin we get
\begin{equation}
f(x) \approx \lambda \left(1+\frac{3}{10} x + O(x^2)\right), \qquad f(0) = \lambda.
\end{equation}
A function like this\footnote{In fact, a one parameter family of functions controlled by $\lambda$.} satisfies, at small angles, the requirement
\begin{equation}
\langle \varphi^i(h) \rangle_{h_0} = \varphi^i(h_0) 
\end{equation}
by construction.

In fact, these operators are much more interesting. Indeed, when we look for the commutation with the fluxes, we can fix $\lambda$ such that
\be\label{cc}
[ R^i , \varphi^j(\coord{h}) ] = R^i \triangleright \varphi^j(\coord{h}) = \delta^{ij} +O(| \coord{h} |^2),
\ee
where $R^i$ is the right invariant vector field on the group.

Expanding \Ref{cc} we get:
\eqa\label{cccond}
R^i \triangleright \varphi^j(\xi_h) &=& \frac{1}{2 i}\frac{d}{\dd s}\, \varphi^j (tr(\vec{\sigma} e^{is\sigma^i} h))|_{s=0}= \frac{1}{2 i} \partial_k \varphi^j(h) tr(\sigma^k i\sigma^i h) = \no
&=&  \partial_k \varphi^j (\delta^{ki} \cos\theta - \epsilon^{ki}{}_l \coord{h}^l). 
\neqa

Using the ansatz \Ref{newcoord} with the result \Ref{sol} , the commutation relation with the fluxes 
\begin{equation}
[R^i,\varphi^j(h)] = \left( f(\coord{h}^2) \delta^{j}_k + 2 f'(\coord{h}^2) \coord{h}^j \coord{h}^k \right)\left( \delta^{ik}\cos{\theta} - \epsilon^{ki}_{\phantom{ki}j }\coord{h}^l \right) \approx \lambda \delta^{ij} + O(\coord{h}^2)
\end{equation}

Consequently, by asking that the holonomies are close to the identity, we put ourselves in the regime where the coordinates $\varphi^i$ are the canonically conjugated variables to the fluxes (choosing $\lambda=1$). 

It is interesting then to look at the behavior of the expectation value of this commutator. The calculation is straightforward but tedious. We only report the result:
\begin{equation}
\la[R^i,\varphi^j(h)]\ra_{h_0} \approx (J_1(t)+J_2(t))\delta^{ij} +\epsilon^{ij}_{\phantom{ij}l } f(\coord{h_0}^2) P_{h_0}^l
\end{equation}
where we recognize the integrals $J_1(t)$ and $J_2(t)$ which we have already introduced. Remembering that we have designed the function $f(x)$ in such a way that
\begin{equation}
J_1(t) + J_2(t) = f(0) = \lambda = 1,
\end{equation}
and that 
the flux is related to the right invariant vector field by
\begin{equation}
\flux^i = i t R^i 
\end{equation}
we conclude that, for small $t,\theta_0$
\begin{equation}
\la [\flux^i,\varphi^j(h)]\ra_{h_0} \approx i t \left( \delta^{ij} 
+\epsilon^{ij}_{\phantom{ij}l } f(\coord{h_0}^2) P_{h_0}^l 
\right)+ O(\theta_0^2) \,.
\label{commu}
\end{equation}

Furthermore, they have the correct expectation values and hence they really represent a serious definition for (approximate) canonically conjugated variables to the
fluxes  (even though only in the regime where the holonomies are close to the identity).
As we have already said, this is not extendible to a statement over the entire phase space, and holds only in certain special circumstances which, nonetheless, are physically relevant.

\subsection{Fluctuations}

Let us now look at the fluctuations. We recall that the fluctuations of both flux and holonomy operators are nicely saturating the uncertainty relations for the Hall states. We start with the fluctuation for the momentum variables. It is enough to compute the fluctuation $\Delta E := \sum_i \langle (\flux^i)^2 -\langle \flux^i \rangle^2 \rangle = \langle \flux^2 \rangle - \langle \flux \rangle^2$. Again working at Fourier space, and under the same assumptions used to compute the expectation value of the flux operator, we have (the $\star$-products are all with respect to the variable $x$):
\eqa
&& \langle \flux^2 \rangle = \bra{\psi^t_{(h_0,y_0)}} \hp^2 \ket{\psi^t_{(h_0,y_0)}} =  \int\dd x \overline{\F(\psi^t_{(h_0,y_0)})}(x) \star x^2 \star \F(\psi^t_{(h_0,y_0)})(x) = \no
&&=  \int\dd x \overline{\F(\psi^t_{h_0})}(x-y_0) \star x^2 \star \F(\psi^t_{h_0})(x-y_0) = \no
&&=  \int\dd x \overline{\F(\psi^t_{h_0})}(x) \star (x+y_0)^2 \star \F(\psi^t_{h_0})(x) = \no 
&&= y_0^2 ||\psi^t_{h_0}||^2 + 2y_0^i \bra{\psi^t_{h_0}} \flux^i \ket{\psi^t_{h_0}} + \bra{\psi^t_{h_0}} \flux^2 \ket{\psi^t_{h_0}}=\nonumber \\[2.5mm]
&&= y_0^2 + \bra{\psi^t_{h_0}} \flux^2 \ket{\psi^t_{h_0}},
\neqa
which implies
\be
\Delta E = \bra{\psi^t_{h_0}} \flux^2 \ket{\psi^t_{h_0}} = \bra{\psi^t_{\1}} \flux^2 \ket{\psi^t_{\1}},
\ee
the same that was found for the state peaked on the identity. We know already \cite{gcs, stw}, then, that the choice of $\psi^t_{h_0}$ as the heat kernel state leads to fluctuations of order $t$, for $t$ small.

\

To compute the fluctuations for the holonomy we need to compute the diagonal matrix element:
\begin{equation}
m^{ij}=\bra{\psi^t_{(h_0,y_0)}} \coord{h}^{i}\coord{h}^{j}\ket{\psi^t_{(h_0,y_0)}}  = \int dh |\psi^{t}_{h_{0}}(h)|^{2} \coord{h}^{i}\coord{h}^{j}=
\int dh |\psi^{t}_{\mathbb{I}}(h)|^{2} \coord{hh_0}^{i}\coord{hh_{0}}^{j}
\end{equation}

Define the following integrals (remember that we are working with normalized states)
\bea
&& I_{1}(t):=\int dh |\psi^{t}_{\mathbb{I}}|^{2} \cos{\theta} = a(t)
\\
&& I_{2}(t):=\int dh |\psi^{t}_{\mathbb{I}}|^{2} \cos^{2}{\theta} 
\\
&& I_{3}(t):= \int dh |\psi^{t}_{\mathbb{I}}|^{2} (\coord{h}^{z})^{2} 
\eea
Remembering that
\begin{equation}
\cos^2{\theta} = {1-\coord{h}^2},
\end{equation}
we find an obvious relation between $I_2$ and $I_3$
\begin{equation}
I_2 = 1-3I_{3}
\end{equation}
These integrals are easily estimated. First of all, from analytic considerations
\begin{equation}
-1< a(t) < 1, \qquad 0 < I_2(t) < 1,
\end{equation}
and 
\begin{equation}
\lim_{t\rightarrow 0} I_1(t) = 1,\qquad \lim_{t\rightarrow 0} I_2(t) = 1,
\end{equation}
since for $t\rightarrow 0$ the heat kernel reduces to a Dirac delta, while
\begin{equation}
\lim_{t\rightarrow \infty} I_1(t) = 0,\qquad \lim_{t\rightarrow \infty} I_2(t) = \frac{1}{4},
\end{equation}
since for large times the heat kernel reduces to a constant function on $SU(2)$. 
\begin{figure}
\framebox{\includegraphics[width=65mm]{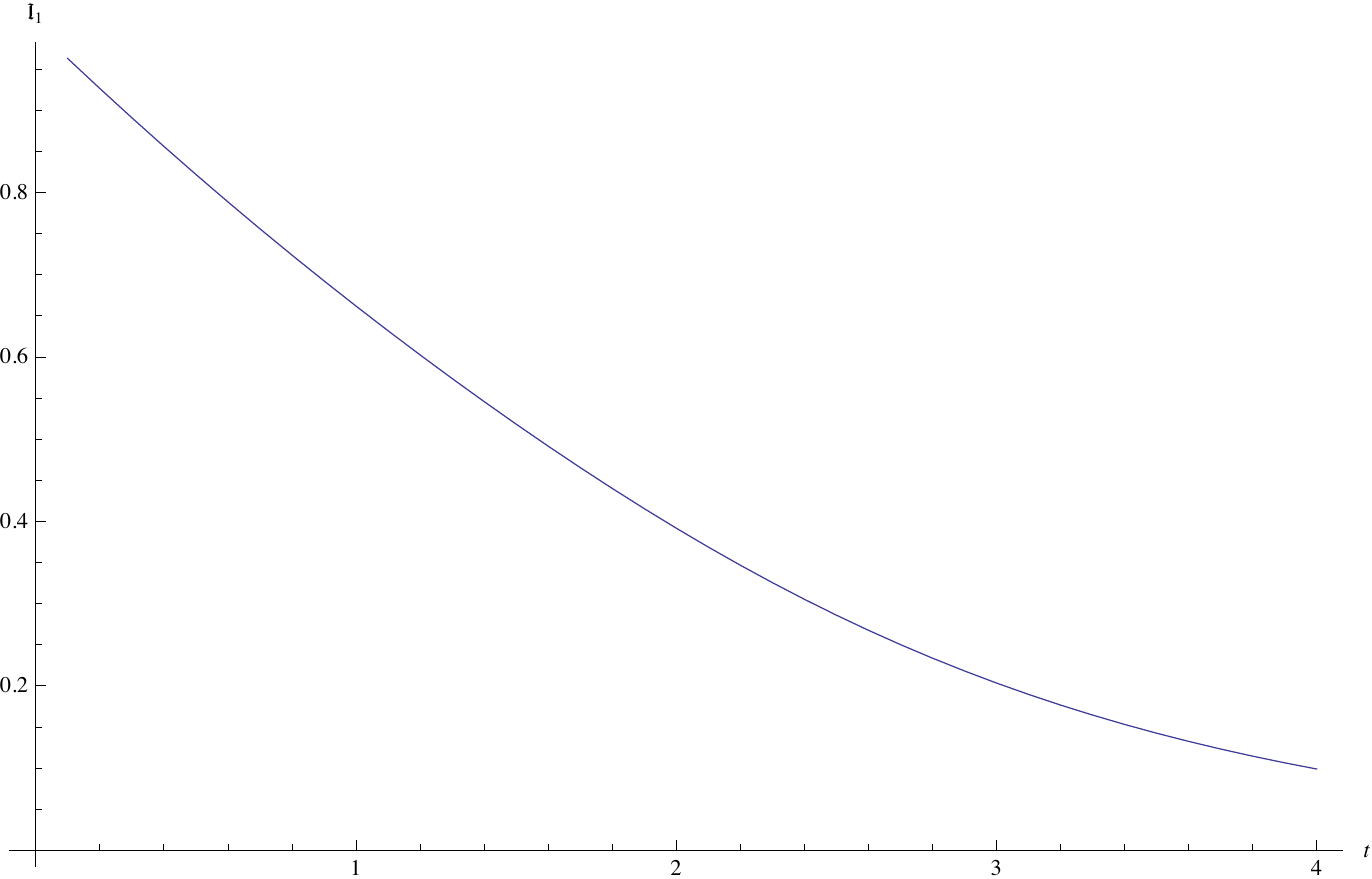}\hspace{1cm}}
\framebox{\includegraphics[width=65mm]{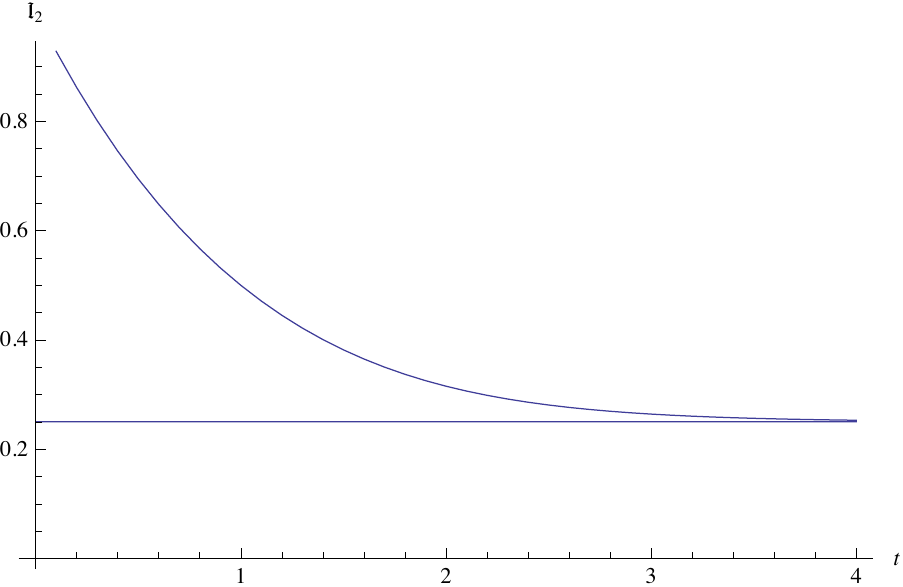}}
\caption{Left panel: $I_1(t)$. Right panel: $I_2(t)$, with the asymptotic value $1/4$ highlighted. }
\label{plots12}
\end{figure}
A numerical 
integration (see Fig. \ref{plots12})  shows that the functions $I_1(t)$ and $I_{2}(t)$ are monotonically decreasing
as $t$ increases, the decrease being essentially exponential and the behavior near $t=0$ being well approximated by a linear function
\begin{equation}
I_{a}(t) \approx 1-\mu_{a} t, \qquad \mu_a = O(1), \qquad a=1,2.
\end{equation}

The desired correlators can be written in terms of these integrals. It is easy to see that, for rotational invariance (but it can be checked with a direct computation)
\begin{equation}
\int dh |\psi^{t}_{\mathbb{I}}|^{2} \coord{h}^{i} \coord{h}^{j} = I_{3}(t) \delta^{ij}
\end{equation}

After straightforward manipulations we obtain
\begin{equation}
m^{ij} = (I_{2}-I_3) \coord{h_0}^i\coord{h_0}^j +  I_3 \delta^{ij}  = \frac{4I_2-1}{3}\coord{h_0}^i\coord{h_0}^j +  \frac{(1-I_2)}{3} \delta^{ij} 
\end{equation}

Given that:
\be
\bra{\psi^t_{(h_0,y_0)}} \coord{h}^{i} \ket{\psi^t_{(h_0,y_0)}} = I_1(t)\coord{h_{0}}^{i}
\ee
we see that the correlations are
\be
\la \coord{h}^{i} \coord{h}^{j} \ra - \la \coord{h}^{i} \ra \la \coord{h}^{j} \ra = \left(\frac{4I_2-1}{3}-I_1^2\right)\coord{h_0}^i\coord{h_0}^j +  \frac{(1-I_2)}{3} \delta^{ij} 
\ee

The dependence on $t$ is hidden in the integrals $I_{1},I_{2},I_{3}$. However, we can say that, for $h_0$ close to the identity, or, equivalently, $\theta_0\ll1$, the correlation can be well approximated by
\begin{equation}
\la \coord{h}^{i} \coord{h}^{j} \ra - \la \coord{h}^{i} \ra \la \coord{h}^{j} \ra \approx   \frac{(1-I_2)}{3} \delta^{ij} + O(\theta_0^2)
\end{equation}
This approximation is valid for any value of $t$. However, if we use the behavior of the function near $t=0$,
\begin{equation}
\la \coord{h}^{i} \coord{h}^{j} \ra - \la \coord{h}^{i} \ra \la \coord{h}^{j} \ra \approx   \frac{\mu_2}{3} t \delta^{ij} + O(\theta_0^2) + O(t^2),
\end{equation}
matching the fact that, for small $t$, the state is well peaked around the mean value.

We might also compute the fluctuations for the coordinates introduced in the previous subsection. 
We will have to deal with the following integral
\begin{equation}
M^{ij}=\int dh |\psi^{t}_{\1}(hh_0)| \varphi^{i}( \coord{h})\varphi^{j}( \coord{h})=
\int dh |\psi^{t}_{\1}(hh_0)| f^2( \coord{h})\coord{h}^i\coord{h}^j.
\end{equation}
Expanding again for small $\theta_0$ we get
\bea
M^{ij} \approx \int dh |\psi^{t}_{\1}(h)| f^2( \coord{h})\coord{h}^i\coord{h}^j + 
\int dh |\psi^{t}_{\1}(h)| f^2( \coord{h})\left(\coord{h}^i \delta\coord{h}^j+\coord{h}^j \delta\coord{h}^i\right)+  \nonumber \\
+4\int dh  |\psi^{t}_{\1}(h)| f( \coord{h})f'( \coord{h})\coord{h}^i\coord{h}^j\coord{h}^k\delta\coord{h}^k \,.
\eea 
Let us consider the term
\begin{equation}
\coord{h}^i \delta\coord{h}^j = \coord{h}^i \left( \cos\theta \coord{h_0}^j - \epsilon_{jrs} \coord{h}^r \coord{h_0}^s \right)\,.
\end{equation}
The first term averages to zero, under integration, while the second gives rise to a contribution of the form
$
H(t) \delta^{ir}\epsilon_{jrs} \coord{h_0}^s
$.
However, this term cancels with the one coming from the evaluation of the integral with $\coord{h}^j \delta\coord{h}^i$. 

Finally, the last term involves:
\begin{equation}
\coord{h}^i\coord{h}^j\coord{h}^k\delta\coord{h}^k = \coord{h}^i\coord{h}^j\coord{h}^k
\left( \cos\theta \coord{h_0}^k - \epsilon_{krs} \coord{h}^r \coord{h_0}^s \right).
\end{equation}
Both of them give zero contribution, for symmetry arguments. Therefore the fluctuation is entirely
determined by the first integral
\begin{equation}
\int dh |\psi^{t}_{\1}(h)| f^2( \coord{h})\coord{h}^i\coord{h}^j = J(t) \delta^{ij},
\qquad
J(t) = \frac{1}{3}\int dh |\psi^{t}_{\1}(h)| f^2( \coord{h})\coord{h}^2 .
\end{equation}
Of course, given that we are truncating at the linear order in $\theta_0$, this is the only contribution to the fluctuation in the new variables $\varphi(\coord{h})$. Again, notice that, for small $t$, this is of order $t$.

\begin{figure}
\framebox{\includegraphics[width=65mm]{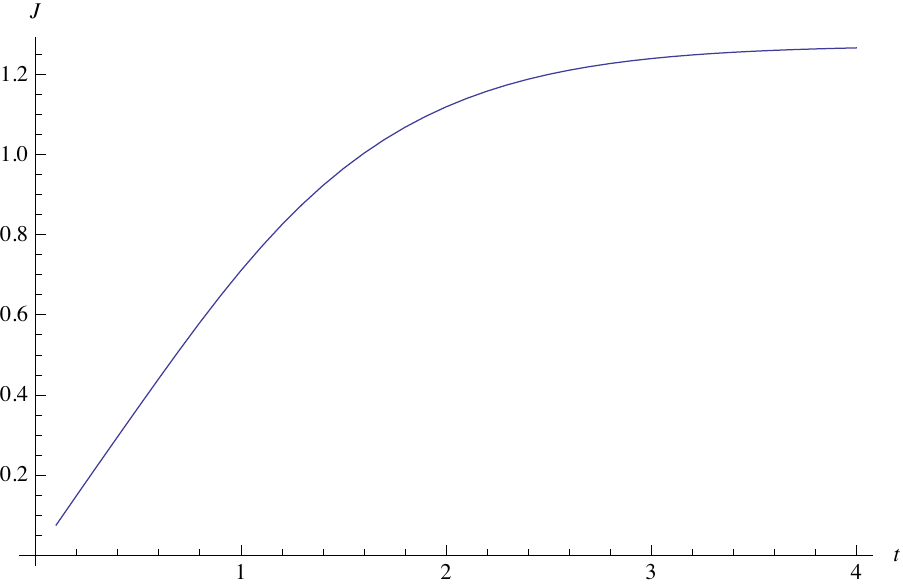}}
\caption{Plot of $J(t)$.}\label{plotj}
\end{figure}
This integral is clearly different from the $I_1,I_2,I_3$ introduced previously  (see Fig. \ref{plotj} for the plot of a numerical estimate).
However, for small
values of $t$ the difference disappears: for small $t$ the Gaussians tend to the common behavior of a Dirac delta, and hence only the domain of integration near the identity gives a significant contribution. In this domain, the difference between the coordinate $\coord{h}$ and the operators
$\varphi(\coord{h})$ tends to zero (of course, if $\lambda=1$), and $J(t)\rightarrow I_3(t)$. Given that, in
this regime, the fluctuation of the new coherent states are identical to those of the Hall states, we
can say that the fluctuations will be minimal as well.

These results collected so far tell us that, for states peaked on holonomies close to the identity (physically relevant situation for low curvature geometries), the statistical properties are essentially determined by the statistical properties of the coherent states constructed with the heat kernel, and hence show that the coherent states introduced here are indeed another (related) class of states
that can be of interest for the investigation of the semiclassical limit of LQG.

\subsection{Resolution of the identity}
We now want to show that the coherent states we have constructed define an (over)complete basis for the space $L^2(G,d\mu(g))$, and thus give a resolution of the identity.

That is, we would like to ask the following condition:
\eqa
\int_{\SU(2)\times \R^3} \mu(h_0,y_0)\; \overline{\psi^t_{(h_0,y_0)}(h)}\,\star_{y_0}\, \psi^t_{(h_0,y_0)}(\tilde{h}) = |\psi^t_{\1}|^2 \delta(h,\tilde{h}), \label{resol}
\neqa
for a certain measure on the classical phase space to be defined. 

Recalling that:
$$ \psi^t_{(h_0,y_0)}(h) \, = \, \sum_\epsilon\, \psi^t_{(h_0,0)}(h) \, e_{h}(y_0)\, I_\epsilon(h) \xi_{\epsilon(h)} $$

and expanding the expression on the l.h.s. of (\ref{resol}), we have (we neglect the characteristic functions from the expression for better simplicity of notation):

\eqa
&&\int_{\SU(2)\times \R^3} \mu(h_0,y_0)\; \overline{\psi^t_{(h_0,y_0)}(h)} \star_{y_0} \psi^t_{(h_0,y_0)}(\tilde{h}) \, = \no
&& = \, \int \mu(h_0,y_0) \;\overline{e_h(-y_0)} \xi_{\epsilon(h)}\,\star_{y_0}\,e_{\tilde{h}}(-y_0)\xi_{\epsilon(\tilde{h})}\, \overline{\psi^t_{(h_0,0)}(h)} \psi^t_{(h_0,0)}(\tilde{h}) =  \no
&& = \int \mu(h_0,y_0) e_{h^{-1}\tilde{h}}(y_0)\,\xi_{\epsilon(h^{-1}\tilde{h})}\, \overline{\psi^t_{(e,0)}(hh_0^{-1})} \psi^t_{(e,0)}(\tilde{h}h_0^{-1}). 
\neqa
where we have assumed once more that $\psi^t_{(h_0,0)}(h) = \psi^t_{(e,0)}(hh_0^{-1})$.

We now make, for the measure on phase space, the natural assumption: $\mu(h_0, y_0) = \dd h_0\, \dd y_0$, with the standard Haar and Lebesgue mesaures on $\SU(2)$ and $\mathbb{R}^3$ respectively.

This gives the desired result:
\eqa
&&\int_{\SU(2)\times \R^3} \mu(h_0,y_0)\; \overline{\psi^t_{(h_0,y_0)}(h)} \star_{y_0} \psi^t_{(h_0,y_0)}(\tilde{h}) \, = \no
&&=\int \mu(h_0,y_0) e_{h^{-1}\tilde{h}}(y_0)\,\xi_{\epsilon(h^{-1}\tilde{h})}\, \overline{\psi^t_{(e,0)}(hh_0^{-1})} \psi^t_{(e,0)}(\tilde{h}h_0^{-1}) = \no 
&& = \int_{\SU(2)}\,\dd h_0\; \delta\left(  h^{-1} \, \tilde{h} \right)\,\overline{\psi^t_{(e,0)}(h h_0^{-1})} \psi^t_{(e,0)}(h h_0^{-1})\, =\, \delta\left(  h^{-1} \, \tilde{h} \right)\, |\psi^t_{(e,0)}|^2\qquad .
\neqa

Note that this result, with the given measure on phase space, holds for any choice of Gaussian and is a direct consequence of the choice of Fourier transform defining the coherent states.

{ It is interesting to compare this result with the corresponding analysis of the resolution of the identity for the Hall states. There, in the case of states on $SU(2)$ analytically continued to $SL(2,\mathbb{C})$, the measure on the  the Lie algebra $\su(2)$ is not simply the Lebesgue measure on $\mathbb{R}^3$ as in this case, but has to be supplemented by an additional factor (see especially the second paper in \cite{gcs}, equation 4.82)
\begin{equation}
\frac{2\sqrt{2}e^{-t/4}}{(2\pi t)^{3/2}} \frac{\sinh{||y||}}{||y||} \exp\left(-\frac{y^2}{2}\right),
\end{equation}
so that the overcompleteness relation can be recovered. The crucial difference between these two cases is that, while in the case of Hall states all the functions involved are seen as ordinary functions on $\mathbb{R}^3$ (isomorphic to the Lie algebra as vector space), the
coherent states that we are proposing here are noncommutative functions on the Lie algebra. 

Incidentally, the integration on the algebra of the star product of two functions can be seen as the
ordinary integration on $\mathbb{R}^3$, provided that the (nonlocal) differential operator $\sqrt{1 + \nabla^2}$ is inserted
\begin{equation}
\int d^3x f(x)\star g(x) = \int d^3x f(x) \sqrt{1+\nabla^2}g(x)
\end{equation}

All these observations also imply that, when seen as ordinary functions on $\mathbb{R}^3$, our coherent states are very different, especially in their asymptotic behaviour, with respect to the (analytically continued) Hall states, also used in LQG\footnote{We would like to thank J. P. Gazeau and an anonymous referee for pointing this out to us.}.

}

\subsection{Overlap}

{ Another important property\footnote{We would like to thank an anonymous referee for this point.} needed for the characterization of the coherent states is the behaviour of the overlap between two states peaked on different phase space points. For this, we need to consider the whether they satisfy

\begin{equation}
\frac{|\langle \psi^t_{(x_1,h_1)}|\psi^t_{(x_2,h_2)\rangle}|^2}{||\psi^t_{(x_1,h_1)}||^2||\psi^t_{(x_2,h_2)}||^2} \approx \left\{ 
\begin{array}{lll}
1 & &(x_1,h_1) \approx (x_2,h_2),\\
&&\\
0 & &(x_1,h_1) \neq (x_2,h_2),
\end{array}
\right.
\label{overlap}
\end{equation}
\ie that, while two coherent states might have a large overlap when the peaks are sufficiently close,
their scalar product becomes smaller and smaller (and, ideally, goes to zero) as the distance between the peaks increases.
  
Before proving the statement, let us remark that the reasoning will hold for any choice of $x_1,x_2$ and, consequently, of $\Delta x= x_2-x_1$. 

As said, we need to understand the behavior of
\begin{equation}
\frac{\left| \int dh\overline{\psi^t_{(x_1,h_1)}(h)}{\psi^t_{(x_2,h_2)}(h)}  \right|^2}{\left(\int dh |\psi^t_{(x_1,h_1)}(h)|^2\right)\left(\int dh |\psi^t_{(x_2,h_2)}(h)|^2\right)}.
\end{equation}
It turns out that, if we construct our states with Hall coherent states on the group, the denominator of this expression is a constant \emph{independent} from $(x_1,h_1)$ and $(x_2,h_2)$. For instance, in considering the norm of the first state, the multiplication of the plane wave with its complex conjugate leads to an expression independent from the Lie algebra element $x_1$,
\begin{equation}
\int dh |\psi^t_{(x_1,h_1)}(h)|^2 = \int dh (\psi^t_{\id}(h h_1^{-1}))^2,
\end{equation}
which, in turn (using the fact that the heat kernel is a class function and that it has a nice behaviour under convolution) leads to
\begin{equation}
\int dh (\psi^t_{\id}(h h_1^{-1}))^2 = \int dh \psi^t_{\id}(h h_1^{-1})\psi^t_{\id}(h_1h^{-1}) = \psi^{2t}_{\id}(\id).
\end{equation}
Consequently, we will need just to address the behaviour of the numerator of the overlap.

Let us start by considering the case in which $h_1,h_2$ are very different (with respect to the scale set by $t$). 
In this case, we can use an estimate for the numerator of \eqref{overlap}.

\begin{equation}
|\int dh \,\overline{e_{h}(x_1)} \overline{\psi^t_{\id}(hh_1^{-1})}
e_{h}(x_2) \psi^t_{\id}(hh_2^{-1}) 
|\leq 2 \int dh {\psi^t_{\id}(hh_1^{-1})}
\psi^t_{\id}(hh_2^{-1}) = \psi^{2t}_{\id}(h_1 h_2^{-1})
\end{equation}
where we have used the fact that the Gaussian that we are using is the heat kernel on $SU(2)$, its reality and the convolution property. Therefore, if $h_1h_2^{-1}$ is sufficiently far away from the identity of $SU(2)$, with respect to the (angular) scale set by the parameter $t$, the decay of the heat kernel is ensuring that the overlap goes to zero as we increase the angle $\theta_{12}$ (measuring the separation on the sphere $S^3$ of the two group elements $h_1$ and $h_2$).

The only nontrivial case, then, is when the two group elements $h_1$ and $h_2$ are similar. In this case, we need to massage the integrals in a different way.
First of all,
\begin{equation}
\int dh \,\overline{e_{h}(x_1)} {\psi^t_{\id}(hh_1^{-1})}
e_{h}(x_2) \psi^t_{\id}(hh_2^{-1}) = \int dh \,{e_{h}(\Delta x)} {\psi^t_{\id}(hh_1^{-1})}
\psi^t_{\id}(hh_2^{-1})
\end{equation}
where $\Delta x=x_2-x_1$. The fact that we are now considering the regime $h_1\approx h_2$ allows us to use the estimate
\begin{equation}
{\psi^t_{\id}(hh_1^{-1})}
\approx
\psi^t_{\id}(hh_2^{-1}).
\end{equation}
Therefore, our integral becomes:
\begin{equation}
\int dh \,{e_{h}(\Delta x)} (\psi^t_{\id}(hh_1^{-1}))^2= \int dh \,{e_{hh_1}(\Delta x)} (\psi^t_{\id}(h))^2 
\end{equation}
after a change of variable of integration. Following the definition
\begin{equation}
e_{hh_1}(\Delta x) = \exp(i P^i(hh_1) \Delta x_i),
\end{equation}
and remembering that
\begin{equation}
P^i(h h_1)= P^i(h)\cos\theta_1 + P^i(h_1) \cos(\theta) - \epsilon^{ijk}P_{j}(h)P_{k}(h_1)
\end{equation}
we obtain
\begin{equation}
e_{hh_1}(\Delta x) = \exp(i P^i(hh_1) \Delta x_i) = \exp(i P(h_1)\cdot \Delta x \cos(\theta))
\exp(i v(\Delta x,h_1)\cdot P(h)),
\end{equation}
where we have introduced the vector
\begin{equation}
v_i(\Delta x,h_1) = \left(\cos\theta_1\delta_{ij} +\epsilon_{ijk} P^{k}(h_1) \right) \Delta x^{j}.
\end{equation}
Notice that it is independent from the variable of integration. At this point we can proceed with the integration, by choosing Euler angles on $S^3$ in such a way that the integral reduces to
\begin{equation}
I=\frac{1}{8\pi}\int d\theta  d\phi d\psi \sin^2\theta \sin \phi  \exp(i P(h_1)\cdot \Delta x \cos(\theta))
\exp(i ||v|| \sin\theta \cos \phi ) F^2_t(\cos(\theta)),
\end{equation}
where we have used the fact that the heat kernel is a class function, $\psi^t_{\id}(h) = F_{t}(\cos(\theta))$. Performing the integration
on the angles $\phi$ and $\psi$ one obtains
\begin{equation}
I=\frac{2}{\pi} \frac{1}{||v||} \int d\theta  \sin\theta \exp(-i P(h_1)\cdot \Delta x \cos(\theta))\sin(||v||\sin\theta)F_t(\cos(\theta))^2.
\end{equation} 
The remaining integral can be bounded, in modulus, by a positive function $f(t)$. Hence, for our purposes, it is enough to examine the prefactor $1/||v||$. From the general expression of the vector $v$, it is clear that, if we assume that $h_1 =\cos\theta_1 \id + i P^i(h_1)\sigma_i$, with $\theta_1\ll1$, 
\begin{equation}
|I| \leq  \frac{f(t)}{||\Delta x||} + O(\sin^2\theta_1),
\end{equation}
where we are neglecting corrections (also depending on $\Delta x$) that would be anyway proportional to $\sin^2\theta_1$, and hence negligible provided that $h_1$ is not too far away from the identity of $SU(2)$. 
Therefore, while this result does not hold for generic $h_1,h_2$ in $SU(2)$, it is valid in the regime in which we are interested, given that it is the only region in which holonomies and fluxes can be approximately be considered as a pair of canonical variables.

In conclusion, the overlap decreases with the inverse of the square of the separation $||\Delta x||$ of the position of the peaks in the Lie algebra. Combining this result with the previous one and with the observation that, for the case in which the two peaks coincide, the overlap is obviously one, we get that the states that we are presenting here are indeed characterized by an overlap that generically decreases with the
increase in the distance between the points in the classical phase space associated to the peaks of the states themselves, at least in the regime in which we are interested in (small holonomies but \emph{generic fluxes}).

Together with the right peakedness properties on phase space, the minimization of the fluctuations of fundamental operators and the resolution of the identity, this last property completes the minimal set of requirements we see fulfilled by our new coherent states based on the flux representation of LQG. }

\section{Conclusion}
We have proposed an alternative definition of coherent states on the cotangent bundle of a compact group, in particular for the $\SU(2)$ case, $\mathcal{T}^*\SU(2)$, of direct interest for quantum gravity. In fact, this work is a contribution to the ongoing, crucial efforts to develop appropriate tools to study the continuum and semi-classical approximation of quantum gravity states defined by (superpositions of) discrete structures labeled by pre-geometric, algebraic data. While the idea behind is more general, if the starting point is a specific choice of Gaussian on the group manifold given by the heat kernel,  it amounts to a simple modification of Hall's construction based on the analytic continuation of group coordinates to peak on generic phase space points. 

Using the non-commutative flux representation for the relevant Hilbert space,  we propose instead to use non-commutative translations on the Lie algebra to achieve the same result. For the new type of coherent states, we have then shown several welcome properties, in particular sharp peakedness with respect to classical phase space points, at least in the very specific regime of holonomies close to the identity. On this point, the improvement with respect to standard Hall states is represented by expectation values that are exactly equal to the classical values in the flux/Lie algebra variable, and that are close to them in a way that is independent of the semi-classicality parameter in the holonomy/group variables (when specific coordinates on the group are chosen).

{Besides the improvement on the expectation values, the fluctuations of the operators around their mean values also display very nice properties. 
We have shown that, for $t$ small, $\Delta \varphi(\coord{h}) \sim t $
and that the relation with the heat kernel states ensures automatically that
$\Delta E  \sim t$, a result that has been established for these special states.  
Putting these results together with the evaluation of the expectation value \eqref{commu} of the commutator
between the fluxes and the coordinates $\varphi$, we see that that the states we have introduced
have fluctuations whose behavior closely resembles the one of ordinary coherent states (in particular, the simultaneous minimality of these
fluctuations), even with respect to the Heisenberg uncertainty relations \begin{equation}(\Delta E\Delta \varphi(P))^{1/2} \sim t \sim |\la [E,\varphi] \ra |.\end{equation} 
In addition to these results concerning the statistical properties (mean values and fluctuations of the basic operators), we add the proof that they define a resolution of the identity and thus an over-complete basis for the Hilbert space, we can conclude that the states that we have introduced here are indeed compelling candidates to be used for the construction of semiclassical states for LQG.

Indeed, in a loop quantum gravity context, the states so-defined correspond to building blocks of generic states associated to graphs, obtained by tensoring them according to the combinatorics of the graph edges and imposing  gauge invariance (Gauss constraint) at the vertices of the graph. The definition of semi-classical states which moreover approximate {\it continuum} phase space points involves then much more than the definition of semi-classical states associated, say, to edges of such graphs, or even to complete graphs. It implies learning to deal with superpositions of graphs or the coarse graining of the same, defining states which are semi-classical with respect to collective observables, rather than  \lq fundamental\rq ones, to define a precise correspondence between discrete and smooth manifolds, and between discrete pre-geometric data and continuum phase space points, even at the classical level, and so on. Many of these issues have been dealt with, to some extent, in the literature \cite{gcs, stw, bombelli}. It is clear, therefore, that our results constitute only a first step toward the stated goal, and a basis for the next steps, that will be the subject of future work.

\section*{Acknowledgements}
This work has been funded by a Sofja Kovalevskaja Award of the A. von Humboldt Foundation, which is gratefully acknowledged. We thank A. Baratin, B. Dittrich, M. Raasakka and J. Tambornino for useful discussions.


\begin{thebibliography}{99}

\bibitem{klauder} Klauder J.R., Skagerstam B-S (1985) Coherent states (World Scientific, Singapore).


\bibitem{lqg}
A. Ashtekar, J. Lewandowski (2004) Background independent quantum gravity: A status report,
 {\em Class Quant Grav} {\bf 21} R53-R152;
  T. Thiemann (2007) \emph{Introduction to Modern Canonical Quantum General Relativity} (CUP);
  C. Rovelli (2004) \emph{Quantum Gravity} (CUP) 
 
\bibitem{stw}
H. Sahlmann, T. Thiemann, O. Winkler (2001) Coherent states for canonical quantum general relativity and the infinite tensor product extension \textit{Nucl.Phys.} \textbf{B606} 401-440, gr-qc/0102038; T. Thiemann, Complexifier coherent states for quantum general relativity, Class.Quant.Grav. 23 (2006) 2063-2118, gr-qc/0206037

\bibitem{BahrThiemann}
B. Bahr, T.Thiemann (2009) Gauge-invariant coherent states for Loop Quantum Gravity. I. Abelian gauge groups \textit{Class.Quant.Grav.} \textbf{26} 045011, arXiv:0709.4619 [gr-qc]; B. Bahr, T.Thiemann (2009) Gauge-invariant coherent states for loop quantum gravity. II. Non-Abelian gauge groups \textit{Class.Quant.Grav.} \textbf{26} 045012, arXiv:0709.4636 [gr-qc]

\bibitem{Hall}
B. C. Hall (1994) The Segal--Bargmann "Coherent State" Transform for Compact Lie Groups, J.Funct.Anal. 122 103

\bibitem{ashetal}
A. Ashtekar, J. Lewandowski, D. Marolf, J. Mourao, T. Thiemann (1996) Coherent state transforms for spaces of connections, J.Funct.Anal. 135 519-551, gr-qc/9412014  

\bibitem{gcs}
T. Thiemann (2001) Gauge field theory coherent states (GCS): 1. General properties, Class.Quant.Grav. 18 2025-2064; T. Thiemann, O. Winkler (2001) 	
Gauge field theory coherent states (GCS). 2. Peakedness properties \textit{Class.Quant.Grav.} \textbf{18} 2561-2636, hep-th/0005237; T. Thiemann, O. Winkler, 	
Gauge field theory coherent states (GCS): 3. Ehrenfest theorems,  Class.Quant.Grav. 18 (2001) 4629-4682, hep-th/0005234; T. Thiemann, O. Winkler, Gauge field theory coherent states (GCS) 4: Infinite tensor product and thermodynamical limit, Class.Quant.Grav. 18 (2001) 4997-5054, hep-th/0005235

\bibitem{bombelli} L. Bombelli, A. Corichi, O. Winkler, Semi-classical quantum gravity: obtaining manifolds from graphs, Class.Quant.Grav. 26 (2009) 245012, arXiv:0905.3492 [gr-qc]; L. Bombelli, A. Corichi, O. Winkler, Semiclassical quantum gravity: Statistics of combinatorial Riemannian geometries, Annalen Phys. 14 (2005) 499-519, gr-qc/0409006 

\bibitem{sf} A. Perez, Spin foam models for quantum gravity, Class.Quant.Grav. 20 (2003) R43, gr-qc/0301113; D. Oriti, Spacetime geometry from algebra: spin foam models for non-perturbative quantum gravity, Rept.Prog.Phys. 64 (2001) 1703-1756, gr-qc/0106091

\bibitem{graviton} E. Bianchi, L. Modesto, C. Rovelli, S. Speziale, Graviton propagator in loop quantum gravity, Class.Quant.Grav. 23 (2006) 6989-7028, arXiv:gr-qc/0604044 [gr-qc]


{E.~Alesci and C.~Rovelli,
``The Complete LQG propagator. I. Difficulties with the Barrett-Crane vertex''
Phys. Rev. D 76, 104012 (2007)
[arXiv:0708.0883 [gr-qc]];

E.~Bianchi, E.~Magliaro and C.~Perini,
``LQG propagator from the new spin foams''
Nucl.\ Phys.\ B 822, 245 (2009)
[arXiv:0905.4082 [gr-qc]].
}

\bibitem{bianchimagliaroperini} E. Bianchi, E. Magliaro, C. Perini, Coherent spin networks, Phys.Rev. D82 (2010) 024012, arXiv:0912.4054 [gr-qc]

\bibitem{acz} A. Ashtekar, A. Corichi, J. A. Zapata, Quantum theory of geometry III: Noncommutativity of Riemannian structures, Class.Quant.Grav. 15 (1998) 2955-2972, gr-qc/9806041

\bibitem{grouptransf}
L.Freidel, E. R. Livine  (2006) Ponzano-Regge model revisited III: Feynman diagrams and effective field theory \textit{Class.Quant.Grav.} \textbf{23} 2021-2062, hep-th/0502106 

\bibitem{karim} 
E. Joung, J. Mourad, K. Noui (2009) Three Dimensional Quantum Geometry and Deformed Poincare Symmetry \textit{J.Math.Phys.} \textbf{50} 052503.

\bibitem{majidfreidel}
L. Freidel, S. Majid (2008) Noncommutative harmonic analysis, sampling theory and the Duflo map in 2+1 quantum gravity \textit{Class.Quant.Grav.} \textbf{25} 045006, hep-th/0601004

\bibitem{maiteflorianetera} M. Dupuis, F. Girelli, E. Livine, Spinors and Voros star-product for Group Field Theory: First Contact, arXiv:1107.5693 [gr-qc]

\bibitem{gft} D. Oriti, The group field theory approach to quantum gravity, in D. Oriti (ed.) {\it Approaches to quantum gravity}, CUP (2009), gr-qc/0607032; D. Oriti, The microscopic structure of quantum space as a group field theory, in G. Ellis, J. Murugan, A. Weltman (eds) {\it Foundations of space and time}, CUP (2011), arXiv:1110.5606 [hep-th]

\bibitem{aristidedaniele} A. Baratin, D. Oriti, Group field theory with non-commutative metric variables, Phys.Rev.Lett. 105 (2010) 221302, arXiv:1002.4723 [hep-th]

\bibitem{aristidedaniele2} A. Baratin, D. Oriti, 	
Quantum simplicial geometry in the group field theory formalism: reconsidering the Barrett-Crane model, to appear in New J. Phys.,  arXiv:1108.1178 [gr-qc]

\bibitem{aristidefloriandaniele} A. Baratin, F. Girelli, D. Oriti, Diffeomorphisms in group field theories, Phys.Rev. D83 (2011) 104051, arXiv:1101.0590 [hep-th]

\bibitem{camporesi} R. Camporesi, Harmonic analysis and propagators on homogeneous spaces, Phys.Rept. 196 (1990) 1-134
 
\bibitem{danielematti} D. Oriti, M. Raasakka, Quantum mechanics on SO(3) via non-commutative dual Variables,  Phys.Rev. D84 (2011) 025003, arXiv:1103.2098 [hep-th]

\bibitem{flux}
A. Baratin, B. Dittrich, D. Oriti, J. Tambornino (2010) Non-commutative flux representation for loop quantum gravity,  Class.Quant.Grav. 28 (2011) 175011, arXiv:1004.3450 [hep-th].
  
\bibitem{perelomov} A. Perelomov, {\it Generalized coherent states and their applications}, Springer (1986)

\bibitem{eterasimone} E. Livine, S. Speziale, A new spin foam vertex for quantum gravity,  Phys.Rev. D76 (2007) 084028, arXiv:0705.0674 [gr-qc]


\end{thebibliography}
\end{document}